%% file: main.tex
\documentclass[
superscriptaddress,
reprint,
 amsmath,amssymb,
 aps,
 pra,
]{revtex4-2}
\usepackage{amsmath}
\usepackage{amsfonts}
\usepackage{amssymb}

\usepackage{graphicx}% Include figure files
\usepackage{dcolumn}% Align table columns on decimal point
\usepackage{bm}% bold math

\usepackage{dsfont}
\usepackage{color}
\usepackage{xcolor}
\usepackage{siunitx}

\usepackage[colorlinks=true,
    linkcolor=blue,
    filecolor=blue,      
    urlcolor=blue,
    citecolor = blue]{hyperref}
\usepackage{braket}
\usepackage{mathtools}
\usepackage[normalem]{ulem}
\usepackage{scalerel}
\usepackage{ bbold }
\usepackage{inputenc}
\usepackage[T1]{fontenc}

\usepackage{tikz}
\usetikzlibrary{quantikz}

\begin{document}

\title{Quantum approximate optimization algorithm for qudit systems} 

\date{\today}

\author{Yannick Deller}
\affiliation{Universit\"at  Heidelberg,  Kirchhoff-Institut  f\"ur  Physik, Im  Neuenheimer  Feld  227,  69120  Heidelberg,  Germany }

\author{Sebastian Schmitt}
\affiliation{Honda Research Institute Europe GmbH, Carl-Legien-Str.\ 30, 63073 Offenbach, Germany}

\author{Maciej Lewenstein}
\affiliation{ ICFO-Institut  de  Ciencies  Fotoniques,  The  Barcelona  Institute  of  Science  and  Technology, Av.   Carl  Friedrich  Gauss  3,  08860  Barcelona,  Spain}
\affiliation{ICREA, Pg. Llu\'is Companys 23, 08010 Barcelona, Spain}

\author{Steve Lenk}
\affiliation{Fraunhofer IOSB, branch Advanced System Technologies IOSB-AST,
Am Vogelherd 90, 98693 Ilmenau, Germany}

\author{Marika Federer}
\affiliation{Fraunhofer IOSB, branch Advanced System Technologies IOSB-AST,
Am Vogelherd 90, 98693 Ilmenau, Germany}

\author{Fred Jendrzejewski}
\affiliation{Universit\"at  Heidelberg,  Kirchhoff-Institut  f\"ur  Physik, Im  Neuenheimer  Feld  227,  69120  Heidelberg,  Germany }

\author{Philipp Hauke}
\affiliation{INO-CNR  BEC  Center  and  Department  of  Physics,University  of  Trento,  Via  Sommarive  14,  I-38123  Trento,  Italy}

\author{Valentin Kasper}
\affiliation{ ICFO-Institut  de  Ciencies  Fotoniques,  The  Barcelona  Institute  of  Science  and  Technology, Av.   Carl  Friedrich  Gauss  3,  08860  Barcelona,  Spain}
\email{valentin.kasper@icfo.eu}

\begin{abstract}
A frequent starting point of quantum computation platforms are two-state quantum systems, i.e., qubits. However, in the context of integer optimization problems, relevant to scheduling optimization and operations research, it is often more resource-efficient to employ quantum systems with more than two basis states, so-called qudits. Here, we discuss the quantum approximate optimization algorithm (QAOA) for qudit systems. We illustrate how the QAOA can be used to formulate a variety of integer optimization problems such as graph coloring problems or electric vehicle (EV) charging optimization. In addition, we comment on the implementation of constraints and describe three methods to include these into a quantum circuit of a QAOA by penalty contributions to the cost Hamiltonian, conditional gates using ancilla qubits, and a dynamical decoupling strategy. Finally, as a showcase of qudit-based QAOA, we present numerical results for a charging optimization problem mapped onto a  max-$k$-graph coloring problem. Our work illustrates the flexibility of qudit systems to solve integer optimization problems.
\end{abstract}
\maketitle

\section{Introduction}\label{Intro}

Integer optimization problems \cite{chenIntegeroptBook09,nemhauserIntOptBookt99} are at the heart of challenging real-world applications, such as scheduling optimization~\cite{IshiharaEVCharging2020}, operations research~\cite{rardinOR2017} and portfolio selection~\cite{CornuejolsFinance2018}.
The practical importance of these problems makes the development of efficient solution algorithms  a particularly active field of research. 
In recent years, quantum  information processing technology has advanced substantially and a multitude of industry-relevant problems have been approached with quantum computing technology, for example with quantum annealing~\cite{hauke2020,yarkoni2021}. Many problems have also been addressed by employing algorithms for gate-based universal quantum computing, such as job-shop scheduling~\cite{venturelliJobShop2016}, 
graph coloring~\cite{marxGraphColoring2004, titiloyeGraphColoring2011,titiloyeGraphColoring-2_2011,ChenDomainWall2021} and flight-gate assignment~\cite{StollenwerkFlightGateAssignement2020,ChenDomainWall2021}.
A paradigmatic example for a hybrid classical-quantum algorithm is the quantum approximate optimization algorithm (QAOA), proposed in Ref.~\cite{FarhiQAOA2014, HadfieldQAOA2019}. 
Further, it was recognized that the QAOA (i) is a
computational model itself~\cite{LloydUniveralQAOA2018}, (ii) can lead to an optimal query complexity~\cite{QAOAQuery2017},
and (iii) exhibits the possibility for quantum advantage~\cite{FarhiSupremacy2016}.
Moreover, important research questions involve the role of quantum effects~\cite{alam2021}, the  choice of the classical optimizer~\cite{lavrijsen2020}, and the performance of the QAOA for low and high depth circuits~\cite{zhou2020,McClean2021}. 

The typical starting point for the QAOA are qubits, i.e., quantum mechanical systems with two basis states. Several qubits can then be used to represent integer numbers. However, such a binary representation of integers can lead to hardware overhead~\cite{Lucas2014,Weggemans2021Solving,ChancellorDomainWall2019,ChenDomainWall2021}, and it may be more resource-friendly to work with quantum systems of a finite basis size with dimension $d>2$, called qudits.
Although, the representation of qudits with arbitrary dimension into elementary qubits is computationally efficient, even small improvements in hardware requirements can be of great practical importance in the era of noisy intermediate-scale quantum (NISQ) devices~\cite{Preskill}.  
In addition, there is an increased interest in employing qudit systems as quantum information platforms~\cite{Wang2020}, and there has been great experimental
progress in realizing quantum information processing with qudits such as photons~\cite{PhotonicQudits}, ions~\cite{IonQudits2021}, superconducting circuits~\cite{SuperconductingQudits},
nuclear magnetic resonance platforms~\cite{NMRQudits}, as well as 
Rydberg atoms~\cite{Weggemans2021Solving,Cohen2021}.

In this article, we discuss the QAOA for qudit systems,
and its possible realization in cold atomic systems with long range interactions, e.g., in cold atomic mixtures~\cite{Kasper2020} or  quantum gases inside an optical cavity~\cite{CavitySystem}.  
Specifically, we elaborate the representations of cost functions and constraints of integer optimization problems with qudits.
Further, we give examples of integer optimization problems such as graph coloring and electric vehicle (EV) charging problems, where the qudit formulation provides a convenient representation of integers.
Finally, we numerically benchmark a simplified charging optimization problem for small instances.

The paper is organized as follows: 
In Sec.~\ref{QAOAQudit}, we discuss the QAOA for qudit systems and how to encode integer cost functions into Hamiltonians employing angular momentum operators and generalized Pauli operators. Further, we discuss different ways to implement linear constraints in qudit systems generalizing the work of Ref.~\cite{HadfieldQAOA2019, HadfieldBoolean2018}.
In Sec.~\ref{Application}, we illustrate the implementation of concrete integer optimization problems and in Sec.~\ref{PerformanceAnalysis}, we numerically analyze the performance of the QAOA for a simplified EV charging optimization problem, which amounts to a graph-$k$-coloring problem with additional coloring cost. 

\section{QAOA for qudit systems}\label{QAOAQudit}
This section revisits the QAOA approach and  discusses how to apply the QAOA to qudit systems. The approach is analogous to the case of qubits, only with enlarged local basis states and operators. We first discuss the Hilbert space for qudits and  operators acting on this Hilbert space, namely angular momentum operators and generalized Pauli operators. These two classes of operators can be implemented experimentally, for example, in atomic mixtures~\cite{Kasper2020} or trapped-ion setups~\cite{IonQudits2021}. Next, we give a summary of the general structure of the QAOA~\cite{FarhiQAOA2014}. Then, we give two different ways of encoding cost functions into Hamiltonians employing angular momentum operators and generalized Pauli operators. The two different encodings may prove advantageous for different experimental qudit implementations. This section mainly provides background information necessary for the following sections.
\begin{figure}[h!]
    \centering
    \includegraphics[width=\columnwidth]{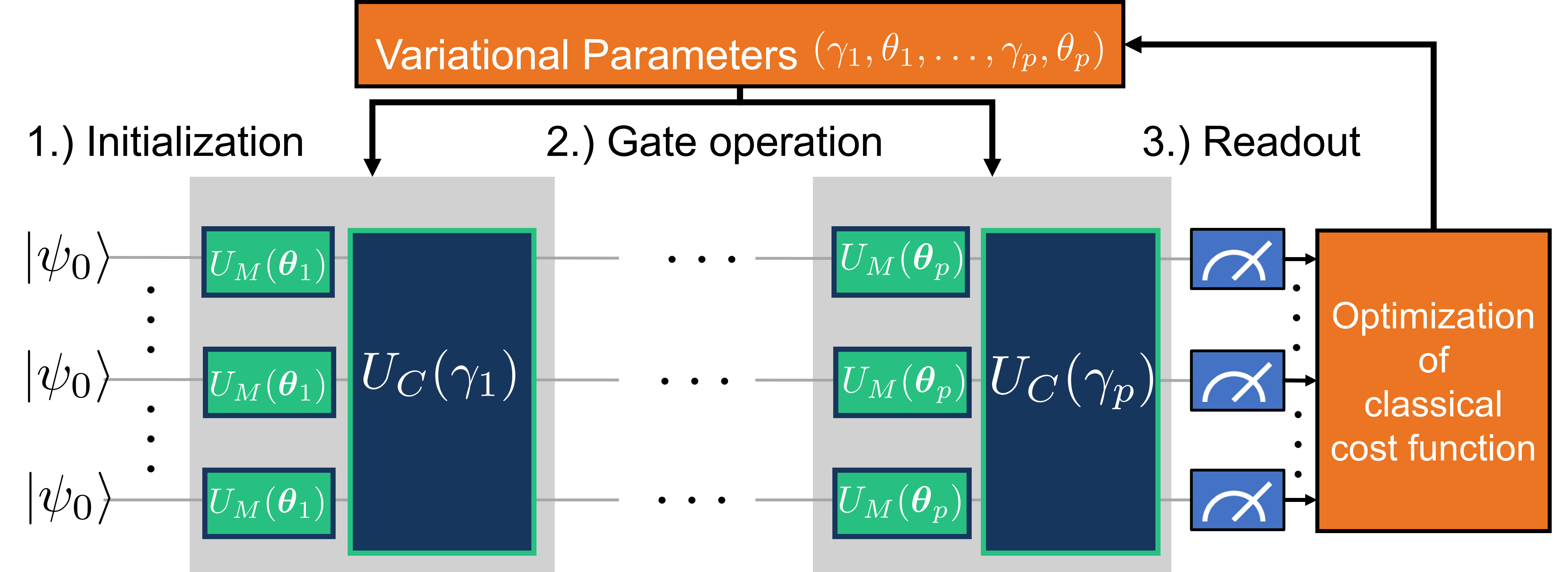}
    \caption{\textbf{QAOA for qudits.} General structure of the QAOA, consisting of preparation of the initial state $\ket{\psi_0}$ which is the equal superposition of all basis states, application of the alternating QAOA-circuit (consisting of phase separation gate $U_C(\boldsymbol{\gamma})$ and mixing gate $U_M(\boldsymbol{\beta})$) and measurement together with subsequent classical optimization of the variational parameters with respect to the expectation value of the cost Hamiltonian.}
    \label{fig:QAOA}
\end{figure}

\subsection{Hilbert space and operators}\label{HilbertSpace}
We consider the $N$-fold tensor product of a $d$-dimensional complex Hilbert space, i.e., $\mathcal{H} = \otimes_{i=1}^N \mathds{C}^{d}$. The total dimension of the Hilbert space is $\text{dim}\, \mathcal{H} = d^N$ and a orthonormal basis for the $d$-dimensional Hilbert space is denoted by $\ket{z}$ with $z \in \{0, \ldots, d-1 \} $. A state vector $\ket{\psi} \in \mathcal{H}$ can be written as
\begin{align}
\ket{\psi} = \sum_{z_1=0}^{d-1} \ldots \sum_{z_N=0}^{d-1}  \alpha_{z_1 \ldots z_N} \ket{\mathbf{z}} \, ,
\end{align}
where $\alpha_{z_1 \ldots z_n}$ is the complex amplitude and the states $\ket{\mathbf{z}} = \ket{z_1, \ldots, z_N }$ form an orthonormal basis, i.e.,  $\braket{\mathbf{z}|\mathbf{z}'}=\delta_{\mathbf{z},\mathbf{z}'}$.

The generalized Pauli $Z$ and $X$-operators~\cite{Wang2020} for one qudit are defined via
\begin{subequations}
\begin{align} \label{eq:Zdef}
    Z &= \sum_{z=0}^{d-1}e^{2\pi i z /d} \ket{z}\!\bra{z} \, , \\
    \label{eq:Xdef}
    X &= \sum_{z=0}^{d-1} \ket{(z+1) \text{mod}\, d}\!\bra{z}  \, ,
\end{align}
\end{subequations}
where the eigenvalues of $Z$ are the roots of unity
\begin{align} \label{eq:ZEVdef}
    Z \ket{z} &= e^{2\pi i z /d} \ket{z} \,.
\end{align}
The definition of the generalized Pauli operators on $\mathcal{H}$ is given by
\begin{subequations}
\begin{align}
Z_j=\underbrace{1\otimes\cdots1\otimes}_{j-1} Z\otimes \underbrace{1\cdots \otimes1}_{N-j} \, , \\
\label{eq:XdefMult}
X_j=\underbrace{1\otimes\cdots1\otimes}_{j-1} X\otimes \underbrace{1\cdots \otimes1}_{N-j} \, ,
\end{align}
\end{subequations}
which only acts non-trivially on the $j^{\mathrm{th}}$ qudit. By construction the basis states $\ket{\mathbf{z}}$ are also eigenstates of products of generalized Pauli $Z$ operators       
\begin{align}
\prod_{j=1}^N Z^{a_j}_{j} \ket{\mathbf{z}} = e^{2\pi i \mathbf{a} \cdot \mathbf{z} /d} \ket{\mathbf{z}}\,,\label{eq:DefGeneralizedPauli} 
\end{align}
where $\mathbf{a} = (a_1, \ldots, a_N)$ with $\mathbf{a} \in\mathbb{Z}^N$ 
summarizes the exponents in the previous expression.

In the following, we define angular momentum operators on the single qudit, which can be realized, for example, in cold atomic gases, see Sec.~\ref{SpinPhononSystems} for details. First, we define the vectors
\begin{align}
    \ket{\ell,m}  \equiv \ket{z}
\end{align}
with $\ell = (d-1)/2$  and $m=z-(d-1)/2$. Using this basis we define the angular momentum operators acting on the local Hilbert space
\begin{subequations}
\begin{align}
    L_z   \ket{\ell,m}   &= m \ket{\ell,m} \, , \\
    L_{+} \ket{\ell,m}   &= \sqrt{(\ell-m)(\ell+m+1)}|\ell,m+1\rangle \, ,\\
    L_{-} \ket{\ell,m}   &= \sqrt{(\ell+m)(\ell-m+1)}|\ell,m-1\rangle\,.
\end{align}
\end{subequations}
Further, the raising and the lowering operators, $L_{+}$ and $L_{-}$, allow us to define the $x$ and $y$ angular momentum operators, 
\begin{subequations}
\begin{align}
    \label{eq:defLx}
    L_x &= \frac{1}{2} (L_+ + L_-)  \,, \\ 
    L_y &= \frac{1}{2i} (L_+ - L_-) \,,
\end{align}
\end{subequations}
which generate rotations around the $x$ and the $y$ axis, respectively.  Finally, we can relate the angular momentum operator to the generalized Pauli $Z$ operator by 
\begin{align} \label{eq:ZwithSz}
Z &=  e^{\tfrac{2\pi i}{d} \big[L_z+(d-1)/2\big]}\, ,
\end{align}
where we used that $\ket{z}$ is an eigenvector of $L_z$.

\subsection{Structure of QAOA}
\label{sec:struct}
The goal of the QAOA is to find the state $\mathbf{z}_0$ which
minimizes a given cost function $C(\mathbf{z})$. In the following, we recount the building blocks and structure of the QAOA~\cite{FarhiQAOA2014, HadfieldQAOA2019}. The starting point of the QAOA is the initial state $\ket{\psi_0}$, for which we assume that it can be prepared efficiently and with high fidelity experimentally. We will frequently use the equal superposition state
\begin{align}
    \ket{\psi_0} = \frac{1}{d^{N/2}} \sum_{\mathbf{z}} \ket{\mathbf{z}},
\end{align}
which is an eigenstate to the generalized Pauli $X$ operators of Eq.~\eqref{eq:XdefMult}. Other initial states are possible, for example an eigenstate to angular momentum operator $L_x$ of Eq.~\eqref{eq:defLx} or even an eigenstate to the $L_z$ operator, e.g.~ $\ket{0,\dots,0}$. We tested several choices of initial states in our numerical experiments and did not find any qualitative differences between these choices. 
It should therefore in principle not matter which initial state is used, and the choice should be guided by which states are most easily prepared in the experimental setup.

The quantum circuit of the QAOA starts from $\ket{\psi_0}$ with subsequent layers of gates (gray boxes in Fig.~\ref{fig:QAOA}). Each layer is composed of two parametrized quantum gates, the so-called phase separation gate $U_C(\gamma)$ and the mixing gate $U_M(\beta)$, which are applied alternatingly. The generator of the phase separation gate is the cost Hamiltonian $H_{C}$ encoding the classical cost function
\begin{equation} \label{eq:phasesep}
    U_C(\gamma)  =  e^{-i \gamma H_C}
\end{equation}
with 
\begin{equation} \label{eq:costHam1}
    H_C \ket{\mathbf{z}} = C(\mathbf{z}) \ket{\mathbf{z}},
\end{equation}
where $\ket{\mathbf{z}}$ denotes the computational basis states.

The mixing gate is defined as
\begin{equation}
    U_M(\beta) = e^{-i \beta H_M}
\end{equation}
 via a mixing Hamiltonian $H_{M}$. Several forms of mixing gates are discussed in the literature~\cite{HadfieldQAOA2019}. Regarding the limitations of the current NISQ-hardware, both phase separation and mixing gate should have an efficient decomposition into the native gate set of the experimental platform, which implements the QAOA.

The QAOA circuit of depth $p \geq 1$ is defined as
\begin{equation}
\label{eq:QAOAUnitary}
U(\boldsymbol{\gamma}, \boldsymbol{\beta})= e^{-i \beta_{p} H_{M}} e^{-i \gamma_{p} H_{C}} \ldots  e^{-i \beta_{1} H_{M}} e^{-i \gamma_{1} H_{C}}\,,
\end{equation}
where $\bm{\gamma}$, $\bm{\beta}\in\mathbb{R}^p$ are free variational parameters to be determined during the execution of the algorithm
and $p$ denotes the numbers of layers.
The trial state 
\begin{align} \label{eq:evolutionQAOA}
    \ket{\bm{\gamma}, \bm{\beta}}&= U(\bm{\gamma}, \bm{\beta})\ket{\psi_0}
\end{align}
is the quantum state approximating a possible solution to the optimization problem. The classical cost function, which is optimized with the QAOA is 
\begin{align} \label{eq:Energy}
 E_{\bm{\gamma},\bm{\beta}}&= \bra{\bm{\gamma}, \bm{\beta}} H_C \ket{\bm{\gamma}, \bm{\beta}} \, ,
\end{align}
which is the expectation value of the cost Hamiltonian.

Typically the variational trial state $ \ket{\bm{\gamma}, \bm{\beta}}$ is a superposition of the computational basis states and the expectation value cannot be obtained in a single experimental run. The expectation value Eq.~\eqref{eq:Energy} is estimated by sampling from the trial wavefunction, see e.g.~\cite{WillschBenchmarkingQAOA2020}. In each sample, a specific configuration $\bm{z}=(z_1,\dots,z_N)$ is obtained from the quantum mechanical trial state with the probability  $\mathcal{P}(\bm{z})=|\!\braket{\bm{z}|\bm{\gamma}, \bm{\beta}}\!|^2$. The expectation value of the cost Hamiltonian is then obtained via
\begin{align}
E_{\bm{\gamma},\bm{\beta}}&\approx \sum_{\text{samples }\bm{z}}  \mathcal{P}(\bm{z}) \,C(\bm{z})\,,
\end{align} 
where $\mathcal{P}(\bm{z})$ is estimated  by sampling from the
final QAOA state $\ket{\bm{\gamma}, \bm{\beta}}$.
In order to obtain a solution to the original optimization problem, one uses a classical optimization method to find the parameters  $\bm{\gamma^*}$ and $\bm{\beta^*}$ that fulfill
\begin{align}
\{\bm{\gamma^*}, \bm{\beta^*}\}&= \underset{\bm{\gamma},\bm{\beta}}{\text{argmin }} E_{\bm{\gamma}, \bm{\beta}}.
\end{align}

After the parameters of the QAOA circuit have been optimized, measuring the output state reveals potential solutions to the optimization problem. In the ideal case, when the QAOA optimization finds an optimal solution, the trial state is a single minimal energy state or a superposition of minimal energy eigenstates of the cost Hamiltonian $H_C$. In particular, when the cost Hamiltonian is invariant with respect to a symmetry transformation and the mixing operator does not break the symmetry, the final state may be a superposition of lowest energy eigenstates. However, frequently minimal energy state cannot be reached, either because the variational trial state cannot faithfully represent the ground state(s) or because the optimization procedure might not find the global optimum. Therefore, the final QAOA state may have contributions from various computational basis states, which are low-energy states and have energies close to the optimal state.

Candidate solutions for the optimization problem are the computational basis states with substantial probabilities. Frequently, the final state needs to be prepared several times in order to sample from the trial state. In an experiment, the selection of the candidate solutions needs to consider the measurement error of the state sampling. In contrast, in numerical studies, the probabilities can be evaluated precisely, and we employ a fixed number of candidate solutions and select the ones with the lowest cost. For analyzing the theoretical performance of the QAOA, we neglect any effect of finite sampling.

\subsection{Cost function Hamiltonian \label{Encoding}}
This subsection discusses two possibilities of mapping certain classical cost functions $C$ to cost Hamiltonians $H_C$. We first discuss a mapping employing generalized Pauli $Z$ operators and then a second mapping using angular momentum operators $L_z$.

\textit{Mapping using generalized Pauli $Z$ operators.} 
The definition of the cost Hamiltonian in Eq.~\eqref{eq:costHam1} implies the following diagonal representation
\begin{align}
    H_C = \sum_{\mathbf{z}} C(\mathbf{z}) \ket{\mathbf{z}}\!\bra{\mathbf{z}} \,.
\end{align}
In order to rewrite the cost Hamiltonian $H_C$ as a polynomial of generalized Pauli $Z$ operators we use the discrete Fourier transform of the cost function, $\widehat{C}$. The Fourier transform and its inverse are given by
\begin{subequations}\label{eq:DFT}
\begin{align}
C(\mathbf{z}) &= \frac{1}{d^N}\sum_{\mathbf{a}} \widehat{C}(\mathbf{a}) e^{2\pi i \mathbf{a} \cdot \mathbf{z}/d}\,,  \\
\widehat{C}(\mathbf{a}) &= \sum_{\mathbf{z}} C(\mathbf{z}) e^{-2\pi i \mathbf{a} \cdot \mathbf{z}/d}\,,
\end{align}
\end{subequations}
with $\mathbf{a} \in \mathbb{Z}^N$ and $0\leq a_j \leq d-1$.
Using the Fourier transform $\widehat{C}$ we can rewrite the cost Hamiltonian as 
\begin{align}
    H_C = \frac{1}{d^N} \sum_{\mathbf{z}}\sum_{\mathbf{a}} \widehat{C}(\mathbf{a}) e^{2\pi i \mathbf{a} \cdot \mathbf{z}/d} \ket{\mathbf{z}}\!\bra{\mathbf{z}} \, .
\end{align}
Employing Eq.~\eqref{eq:DefGeneralizedPauli} the Hamiltonian becomes
\begin{align}
H_{C} &=\frac{1}{d^N} \sum_{\mathbf{a}} \widehat{C}(\mathbf{a})  \prod_{j=1 }^N Z^{a_j}_{j} \,,
\label{eq:costHamGenZ}
\end{align} 
which is a polynomial in the generalized Pauli $Z$ operators~\cite{HadfieldBoolean2018}. This encoding is especially useful when the Fourier transform of the cost function has few nonzero Fourier coefficients. 

\textit{Mapping using $L_z$ operators.}
Here we focus on polynomial functions $C(\mathbf{z})$ in the variables $z_m$. We obtain the cost  Hamiltonian by substituting $z_m \rightarrow (\ell + L_{z,m})$ into the cost function leading to 
\begin{align}
    H_C &= C\Big( \ell + L_{z,1}, \ldots, \ell + L_{z,n} \Big) \,. \label{eq:CostHamLz}
\end{align}
For examples and representations of cost functions we refer to Sec.~\ref{Application} and Appendix~\ref{appendix}.

\subsection{Mixing Hamiltonian \label{Mixing}}
The mixing operator has to be able to traverse the allowed state space of the optimization problem, see  Ref.~\cite{HadfieldQAOA2019}. 
For the local qudit Hilbert space with $d$ levels, $d^2-1$ local operators are in principle necessary to form an operator basis. 
However, as shown previously \cite{Kasper2020,giordaQuditAlgebra2003} a reduced set of three operators is sufficient to generate any state by (possibly many) repeated finite rotations.
In this work, the cost Hamiltonians include linear and higher order terms in $L_z$, which allows us to consider a  mixing Hamiltonian  based only on the angular momentum operator in the $x$ direction
\begin{align} \label{eq:StandardMixer}
    H_M=\sum_{i=1}^N L_{x,i} \, ,
\end{align}
which fulfills the above mentioned criteria for unconstrained integer optimization problems. As detailed in appendix \ref{SpinPhononSystems}, this mixing operator can be experimentally implemented in atomic qudit systems. 
Another viable choice would be to use the generalized Pauli $X$ operators of Eq.~\eqref{eq:XdefMult} as basis for the mixing Hamiltonian.
In principle this should not make a  qualitative difference, which we explicitly confirmed by testing both choices in our numerical experiments.

\subsection{Constraints}\label{Constraints}
In many important optimization problems, the variables of the cost function must satisfy constraints, which can be given by equalities or inequalities, i.e., 
\begin{align}\label{LinearConstraints}
   g_m(\bm{z}) \le 0 \text{ or }  g_m(\bm{z})  = 0
\end{align}
with $m = 1, \ldots, M$. Hence, it is an important question how to incorporate constraints in the QAOA. One common way to enforce constraints into QAOA circuits is by adding appropriate penalty terms to the cost function. Alternatively, one can engineer the mixing operator such that the evolution of the quantum state only takes place in the space of feasible solutions~\cite{HadfieldQAOA2019}. In this subsection, we explicitly implement strategies to enforce constraints in the QAOA. The results developed here are applicable to both qudit and qubit systems. 

A standard route to implement constraints in classical optimization is by adding penalty terms to the cost function
\begin{align}
\tilde C(\mathbf{z})=C(\mathbf{z})+\sum_m \lambda_{m} P_m[g_{m}(\mathbf{z})]\,, \label{eq:CostFunctionWithPenalty}
\end{align}
where $\lambda_m$ are the penalty factors, $P_m$ are the penalty functions, and $g_m(\bm{z})$ are the constraints given by Eq. \eqref{LinearConstraints}. Adding penalty terms is typical for black-box optimization~\cite{MezuraConstraintNatureOpt2011,smithPenalty97,Michalewicz95asurvey}. Possible penalty functions \footnote{Other penalty functions are possible, but they have to fulfill the minimal requirement to (a) produce the same constant value (typically zero) for  all solutions which fulfill the constraint, (b) to be larger than the constant value for infeasible solutions, and (c) to increase monotonically with the degree of constraints violation.} are
\begin{align}
\label{eq:EqualPenalty}
P_\text{eq}\left[g(\bm{z})\right]&=|g(\bm{z})|^a
\end{align}
for equality constraints $g(\bm{z})=0$, and 
\begin{align}
\label{eq:InequalityPenalty}
P_\text{ineq}\left[ g(\bm{z}) \right]&=\text{max}\left[0,g(\bm{z}) \right]^a 
\end{align} 
for inequality constraints $g(\bm{z})\leq 0$, with typical values for the exponents being $a=1$ and $a=2$. One way to implement the modified cost function $\tilde C(\mathbf{z})$ in the QAOA is to use the cost Hamiltonian $H_C$ generated by $C(\mathbf{z})$ for the quantum circuit, but employ $\tilde C(\mathbf{z})$ in the classical optimization loop with the cost function
\begin{align}
    \tilde{E}_{\bm{\gamma},\bm{\beta}}&=\sum_{\bm{z}}  |\braket{\bm{z}|\bm{\gamma}, \bm{\beta}}|^2 
      \left\{ C(\bm{z}) +\sum_m \lambda_{m} P_m\left[g_{m}(\mathbf{z})\right]\right\}  \,.
\end{align}

In addition, one has to tune the penalty parameters during the optimization process. Not including the constraints into the cost Hamiltonian simplifies the experimental realization of the phase separation gate and relaxes the requirements on the hardware. However, depending on the optimization problem and, in particular, on how the constraints confine the feasible search space, the trial states may produce infeasible solutions for randomly chosen $\boldsymbol{\gamma}$ and $\boldsymbol{\beta}$.

In a situation where the cost Hamiltonian and the mixer are invariant with respect to a symmetry, the trial state is also symmetric. 
However, if one constraint violates the symmetry (not implemented in the Hamiltonian), the trial state will not be able to reflect this violation of the symmetry. This scenario may render the QAOA less effective because more candidate solutions must be sampled.

An alternative is to implement the constraint in the circuits. Here, we present three different ways to implement constraints into the quantum circuit: (i) including penalty terms in the cost Hamiltonian, (ii) using conditional gates, and (iii) employing dynamical decoupling.
\subsubsection{Penalty terms in the cost Hamiltonian}
\label{sec:eqConstraint}
One can include constraints into the QAOA by using $H_{\tilde{C}}$ instead of $H_C$, see~\cite{hauke2020,yarkoni2021} for details. 
The penalized cost Hamiltonian is given by
\begin{align}
    H_{\tilde{C}} & = H_C + \sum_m \lambda_m P_m(G_m) \, ,
\end{align}
where we introduced the constraint operator $G_m$ via
\begin{align}
    G_m \ket{\mathbf{z}} = g_m(\mathbf{z}) \ket{\mathbf{z}}\,.
\end{align}
The cost Hamiltonian including the constraints also has a diagonal 
representation
\begin{align}
    H_{\tilde{C}} & = \sum_{\mathbf{z}} \tilde{C}(\mathbf{z}) \ket{\mathbf{z}}\!\bra{\mathbf{z}}\, ,
\end{align}
where $\tilde{C}(\mathbf{z})$ is given by Eq.~\eqref{eq:CostFunctionWithPenalty}. 
Using the results of Sec.~\ref{Encoding} one can write the penalized Hamiltonian $H_{\tilde{C}}$ with the help of generalized Pauli $Z$ or angular momentum operators, where the angular momentum encoding works when $\tilde{C}(\mathbf{z})$ is a polynomial. Again, constructing the Hamiltonian $H_{\tilde{C}}$ with generalized Pauli operators involves a discrete Fourier transform, see Eq.~\eqref{eq:DFT}, which results in polynomials of generalized Pauli $Z$ operators. Here, the $\text{max}$ function in the inequality constraint may introduce higher powers of generalized $Z$ operators. 

Using the cost Hamiltonian $H_{\tilde{C}}$ in the unitary evolution of Eq.~\eqref{eq:QAOAUnitary} leads to variational states $\ket{\bm{\gamma},\bm{\beta}}$ that satisfy the  constraints $g_m$ for appropriate choice of the penalty parameters $\lambda_m$.  A major disadvantage of including penalties in the cost Hamiltonian is the necessity for tuning the penalty factors $\lambda_m$. The penalty factors directly affect the cost function landscape and thus significantly influence search performance.

\subsubsection{Constraints via conditional gates}\label{ConditionalGatesInequalities}

\begin{figure}
    \centering
    \includegraphics[width=\columnwidth]{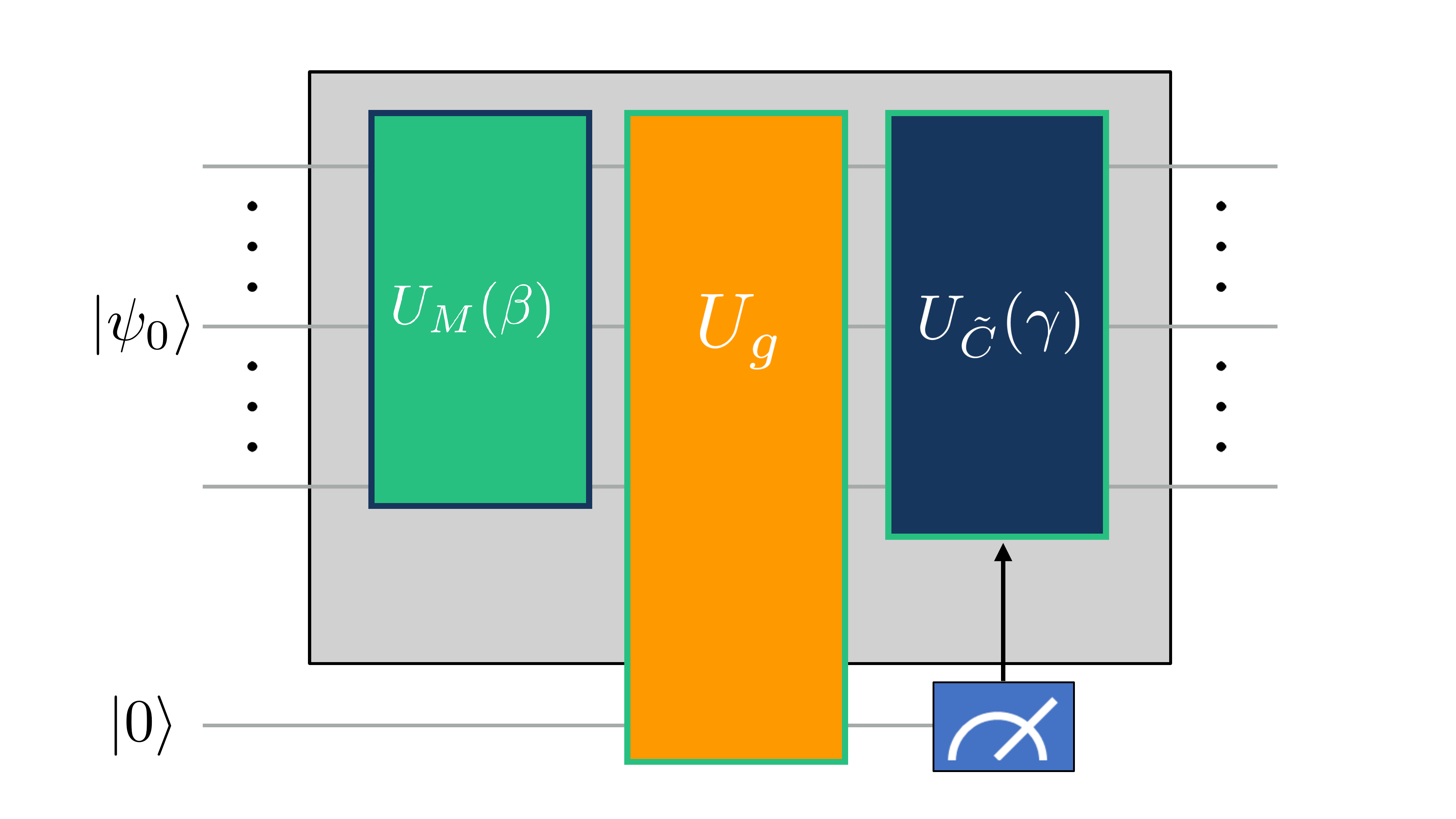}
    \caption{\textbf{QAOA with conditional gates for constraints:} In order to enforce constraints 
    one performs a conditional gate $U_g$ and measures the ancilla qubit $\ket{0}$. If the measurement returns the value $0$, the constraint is fulfilled and we do not change the 
    cost function. If the measurement yields the value $1$, then the constraint is violated and we forward the measurement result into the cost function and use the Hamiltonian $H_{\tilde{C}}$ and penalize the violation of the constraint.}
    \label{fig:PunishmentTerms}
\end{figure}

Here we enforce equality and inequality constraints via conditional gates~\cite{HadfieldBoolean2018}. Specifically, we introduce the unitary operator $U_{g}$ that acts on the $N$ computational qudits and on one ancilla qubit $\ket{y} = \alpha \ket{0} + \beta \ket{1}$ via 
\begin{align}
    U_{g} \ket{\mathbf{z}}\!\ket{y} =  \begin{cases}
    \ket{\mathbf{z}}\!\ket{y}       & \text{for } g(\mathbf{z}) = 0 \text{ or } g(\mathbf{z}) \leq 0\,,  \\
    \ket{\mathbf{z}} X\ket{y} & \text{otherwise }\,.
    \end{cases} \label{eq:ConditionalXGate}
\end{align}
If the quantum state $\ket{\mathbf{z}}$ does fulfill the constraint the ancilla qubit does not change, whereas we apply $X$ on the ancilla, if the constraint is not fulfilled. 
The operator $U_{g}$ belongs to the class of conditional operators:
\begin{subequations}
\begin{align} \label{eq:UQProj}
U_{g}(Q) =\sum_{\mathbf{z}: g(\mathbf{z}) \leq 0}\ket{\mathbf{z}}
\!\bra{\mathbf{z}}\otimes I+\sum_{\mathbf{z}: g(\mathbf{z}) > 0 } \ket{\mathbf{z}}
\!\bra{\mathbf{z} } \otimes Q \,, \\
U_{g}(Q) =\sum_{\mathbf{z}: g(\mathbf{z}) = 0}\ket{\mathbf{z}}
\!\bra{\mathbf{z}}\otimes I+\sum_{\mathbf{z}: g(\mathbf{z}) \neq 0 } \ket{\mathbf{z}}
\!\bra{\mathbf{z} } \otimes Q \,, 
\end{align}
\end{subequations}
where the first line is for inequalities and the second line for equalities. Further, $I$ is the identity operator and $Q$ is an arbitrary unitary operator. The conditional operator $U_g(Q)$ does only apply
the operator $Q$ on the ancilla qubit if the quantum state $\ket{\mathbf{z}}$ does not fulfill the inequality $g(\mathbf{z}) \leq 0$ or equality $g(\mathbf{z}) = 0$. 

We represent $U_g(Q)$ as a matrix exponential in order to discuss the implementation with an appropriately chosen Hamiltonian. Direct calculation~\cite{HadfieldBoolean2018} shows 
\begin{align}
U_g(Q)=e^{-i H_{ g} \otimes H_{Q}} \,,     
\end{align}
with 
\begin{subequations}
\begin{align}\label{ProjectorH}
H_{ g}&=\sum_{\mathbf{z}}  \left[ 1 - \delta_{\text{max} \left[ g(\mathbf{z}),0)\right],0 } \right] |\mathbf{z}\rangle\langle\mathbf{z}|\,, \\
H_{ g}&=\sum_{\mathbf{z}}  \left[ 1-\delta_{g(\mathbf{z}),0} \right] |\mathbf{z}\rangle\langle\mathbf{z}| \,
\end{align}
\end{subequations}
and  
\begin{align}
    H_Q= (\pi /2 ) X \,.
\end{align}
Particularly, we can implement $H_{g}$ again using the Pauli operator encoding.
In order to enforce one constraint $g$ after one layer of the QAOA circuit, we initialize the ancilla qubit in $\ket{0}$ and apply the conditional gate Eq.~\eqref{eq:ConditionalXGate} on the quantum state $\ket{\psi}$, which we assume fulfills all constraints, leading to
\begin{align}
U_g  e^{-i \beta H_{M}} \ket{\psi}\ket{0} \, 
\end{align}
and measure the ancilla qubit $\ket{y}$. If the measurement on the ancilla qubit returns the value zero, the constraint is fulfilled, and we apply the unconstrained cost function Hamiltonian $H_C$. On the other hand, if the measurement of the ancilla qubit yields the result one, the inequality is violated, and we use $H_{\tilde{C}}$ and the mixing Hamiltonian in the next step. Then we apply the conditional unitary operator $U_g$ again and iterate this procedure. 
The quantum circuit illustrating this approach is schematically shown in  Fig.~\ref{fig:PunishmentTerms}.

Using conditional gates is more involved as it requires an additional qubit for tracking the constraint. Finally, conditional gates as described above effectively implement the $\text{max}$ function in the cost Hamiltonian for inequality constraints.

\subsubsection{Equality constraints via dynamical decoupling}
Another way to implement equality constraints is via dynamical decoupling~\cite{Viola1998, Viola1999}, a quantum control technique suppressing coupling to an environment. The technique dates back to nuclear magnetic resonance experiments~\cite{Hahn1950, Carr1954, Meiboom1958}, for a review see~\cite{Lidar2012}. Dynamical decoupling techniques can also be employed to suppress transitions to undesired subspaces, see for example~\cite{Halimeh2020,KasperDD} in the context of quantum simulations of lattice gauge theories. Here we discuss how to use 
dynamical decoupling to enforce constraints. 

To apply dynamical decoupling to the QAOA, we choose an initial state $\ket{\psi_0}$ that already fulfills all equality constraints
\begin{align} \label{eq:constraintInit}
G_m \ket{\psi_0}= {0}
\end{align}
for all $m$.  The unitary operator $U(\boldsymbol{\gamma}, \boldsymbol{\beta})$  of the QAOA may lead to a trial state  $\ket{\boldsymbol{\gamma},\boldsymbol{\beta}}$ where the constraints are not fulfilled, i.e.\,, 
\begin{align} \label{eq:constraintU}
G_m \ket{\boldsymbol{\gamma},\boldsymbol{\beta}} \neq {0} \,.
\end{align}
The goal is to construct a unitary mixing operator that does not evolve the initial state out of the feasible subspace given by the equality 
constraints. Therefore, the mixing term must commute with all constraints. In order to obtain such a unitary mixing term, we employ a dynamical decoupling strategy. 

For simplicity, we assume that $G_m$ has only integer eigenvalues and denote the largest eigenvalue by $\Lambda_m$. We start from the identity  
\begin{align}
    e^{-i\theta G_m}  \ket{\psi_0}&= \ket{\psi_0}\;&\forall\: \theta \in \mathds{R} \, ,
\end{align}
which follows from Eq.~\eqref{eq:constraintInit}. We
define the symmetrization operation of any operator $\mathcal{O}$ by
\begin{align} 
\label{eq:symmetrization}
\bar{\mathcal{O}} =  \prod_m \int_{0}^{\frac{2\pi}{\Lambda_m}} \frac{d\theta_m}{(2\pi/\Lambda_m)}  e^{-i\theta_m G_m} \mathcal{O} e^{i\theta_m G_m}\, .
\end{align}
Specifically, the symmetrization implies 
\begin{align}
e^{-i\phi G_m} \bar{\mathcal{O}} e^{i\phi G_m} = \bar{\mathcal{O}}
\end{align}
for all $m$, which follows from using the integer spectrum of $G_m$ and shifting the integration variables. The above equation is equivalent to $[ e^{-i\phi G_m}, \bar{\mathcal{O}}] = 0 $ for all $m$.
Equality constraints with rational spectrum can always be reformulated as constraints with integer spectrum by multiplying with the least common multiple, while constraints with an irrational spectrum can be approximated with a rational spectrum. 
Employing Eq.\eqref{eq:symmetrization} we symmetrize the mixing Hamiltonian according to  
\begin{align}\label{DynamicalDecouplingMixing}
\bar{H}_M  = \prod_m \int_{0}^{\frac{2\pi}{\Lambda_m}} \frac{d\theta_m}{(2\pi/\Lambda_m)} e^{-i\theta_m G_m}  H_M e^{i\theta_m G_m}\,,
\end{align}
which leads to $[ e^{-i\phi G_m}, \bar{H}_M   ]  =0$  for all $m$ and its infinitesimal version  
$[ G_m, \bar{H}_M   ]  =0 $ for all $m$. Using the symmetrized  mixing Hamiltonian we engineer a new unitary mixing operator
\begin{align}
    {U}_M(\beta) = e^{i\beta \bar{H}_M }
\end{align}
that does not commute with the cost Hamiltonian~\cite{HadfieldQAOA2019} but
guarantees that the final state 
\begin{align}
G_m U(\boldsymbol{\gamma}, \boldsymbol{\beta})\ket{\psi_0}=0\,,
\end{align}
also fulfills the constraints given $G_m \ket{\psi_0}=0$.

Implementing the continuous integral of the the dynamical decoupling strategy of Eq.~\eqref{eq:symmetrization} in a circuit is challenging. 
One strategy is to sample or discretize the integral into a finite sum and use Floquet engineering to determine an appropriate discretization, see, e.g., \cite{Lidar2012}. 
Because of this sampling at each
layer, the dynamical decoupling strategy is only possible for low 
circuit depth. However, in order to reduce the number of decoupling
layers one can selectively introduce the dynamical decoupling,  especially in the last layer.

\section{Applications}\label{Application}
This section discusses optimization problems involving integer variables, which can be addressed with the QAOA based on qudits. Primarily, we illustrate the encodings of section~\ref{Encoding}, which leads to feasible implementations in current qudit systems. Specifically, we treat a graph coloring problem and the optimization of an electric vehicle charging plan. Further integer optimization problems, i.e., a knapsack problem, multiway number partitioning, job-shop scheduling, and their respective qudit encodings, can be found in Appendix~\ref{appendix}.

\subsection{Graph coloring} \label{sec:graphColor}
Let $G=(V, E)$ be a graph with $N$ vertices and $M$ edges. 
A proper vertex $k$-coloring of $G$ is given, if one can assign one of $k$ colors to each vertex such that adjacent vertices have different colors. If one can find such a proper vertex $k$-coloring, the graph $G$ is $k$-colorable~\cite{Lucas2014, HadfieldQAOA2019}. We denote the assignment of colors to the vertices by $\mathbf{z}=(z_{1}, \ldots, z_{N})$ with $z_{i} \in \{0, \ldots, k-1 \}$. The coloring task can be expressed as finding the minimum of an objective function which counts the number of edges between nodes with the same color, i.e.,
\begin{align}
    C(\mathbf{z})=\sum_{(n, m) \in E} \mathrm{\delta}_{z_{n}, z_{m}}\,,
\end{align}
where $n$ and $m$ denote vertices of the graph, $z_n$ and $z_m$ denote the color of the vertices, and $E$ is the set of edges of the graph.

Minimizing $C(\mathbf z)$ leads to the largest induced subgraph that can be properly $k$-colored. Moreover, the cost function can be encoded with Pauli $Z$ operators~\cite{bravyi2020} into the cost Hamiltonian
\begin{align}
H_C&=\frac{1}{k^{N-1}}\sum_{(n, m) \in E} \sum_{a,b=0}^{k-1}  \delta\big((a+b) \text{ mod } k ,0\big)  Z_n^a Z_m^{b} \nonumber\\
&=\frac{1}{k^{N-1}}\sum_{(n, m) \in E}   \left( 1 +  \sum_{a=0}^{k-1} Z_n^a Z_m^{k-a} \right)\, , 
\end{align}
where we used the Fourier transform of the Kronecker-delta, $\widehat{\delta}(a,b)=k \delta[(a+b) \text{ mod } k ,0]$.
For $k=3$ this becomes
\begin{align}
H_C&=\frac{1}{3^{N-1}}\sum_{(n, m) \in E} \left(1+ Z_m^3 + Z_n Z_m^2+ Z_n^2 Z_m \right) \, .
\end{align}
This expression can be reformulated in terms of angular momentum operators $L_{z,n}$ resulting in a polynomial in powers of $L_{z,n}$ and $L_{z,m}$ with $L_{z,m}^{d-1} L_{z,n}^{d-1}$ as the largest power. 

\subsection{Charging optimization}
Many problems in the energy management domain require optimizing a schedule for the distribution of electrical energy among technical devices. A representative problem is the charging schedule of electric vehicles (EVs). Designing these schedules typically leads to integer or mixed-integer programming problems, see for example \cite{vujanicMILP_EVCharging2016, IshiharaEVCharging2020,maoEVCharging2019,hanEVCharging2017}. We consider the following EV charging problem: An operator of charging stations needs to charge $N$ EVs during the working hours of a business complex. The operator can purchase and sell energy for real-time electricity market prices and charge/discharge each EV. The goal is to minimize the electricity cost for the operator while meeting operation constraints. Possible constraints are: (i) Each EV has a desired target state of charge (SOC) of the battery, which needs to be reached at the end of the time period. (ii) Each battery has a minimal and maximal SOC. (iii) There is an upper and lower limit for the total cumulative charging power of all vehicles at all times. 
A simple variant of the EV charging problem without constraints and preemptive charging was discussed in Ref.~\cite{dalyacSmartChargingQAOA2021} using the QAOA for qubits.

The charging plan for the EVs must be optimized for the total time $T$, where we divide $T$ into equidistant time steps of duration $\Delta t$. In each time step the EV $n$ can either be charged, not charged or discharged, which is represented by the ternary variables $L_{n,t} = 1$, $L_{n,t} = 0$ or $L_{n,t} = -1$ respectively. The total electricity cost function is 
\begin{align} \label{eq:EVcost}
C(\mathbf{L})= \sum_{n=1}^N\sum_{t=1}^T\Delta t P^0_n \left(\tfrac{e^c_t+e^d_t}{2}\,L_{n,t}+\tfrac{e^c_t-e^d_t}{2}\,L^2_{n,t}\right)\, ,
\end{align}
where $P^0_n$ is the charging and discharging power of vehicle $n$, and the prices for buying and selling energy are $e^c_t$ and $e^d_t$. Specifically, the charging costs for car $n$ are $\Delta t P^0_n e^c_t$, whereas the discharging costs are $-\Delta t P^0_n e^d_t$.

The SOC of each battery at time $t$ is
\begin{align}
E_{n,t}=E_n^\mathrm{init}+\sum_{k = 1}^t\Delta t P^{0}_n (  L_{n,k} - \delta_n\, L_{n,k}^2) \,,
\end{align}
where $E_n^\mathrm{init}$ denotes the initial SOC of EV $n$, and $\delta_n\geq0$ encodes conversion losses in the EV since the SOC increases (decreases) by $(\pm1-\delta_n) \Delta t P_n^0$ during charging (discharging) with power $P^0_n$. The constraints on the SOC of each battery are
\begin{subequations}
\begin{align} \label{eq:EVSOCconstraint}
E_{n,T}&\geq E_n^\mathrm{target} &\forall n\,, \\
E_n^\mathrm{min} &\leq E_{n,t}   &\forall n,t\,, \\
E_{n,t} &\leq E_n^\mathrm{max}   &\forall n,t \, ,
\end{align}
\end{subequations}
where $E_n^\mathrm{target}$ is the required minimal final SOC at time $t=T$, and $E_n^\mathrm{min}$ and $E_n^\mathrm{max}$ specify the generally allowed SOC for vehicle $n$. The limits on the maximum charging and discharging power are 
\begin{subequations}
\begin{align} \label{eq:EVPconstraint}
P^\mathrm{min} &\leq \sum_n P^{0}_n L_{n,t}   \quad&\forall t \,, \\
P^\mathrm{max} &\geq \sum_n P^{0}_n L_{n,t}    \quad&\forall t  \,,
\end{align}
\end{subequations}
with $P^\mathrm{min}<0$ the largest possible discharging power and $P^\mathrm{max}>0$ the maximum charging power. This amounts to $N(1+2T)+2T$ constraints in total.

\subsection{Combination of charging and graph coloring}
\label{sec:simpSched}
\begin{figure}
    \centering
     \includegraphics[width=\columnwidth]{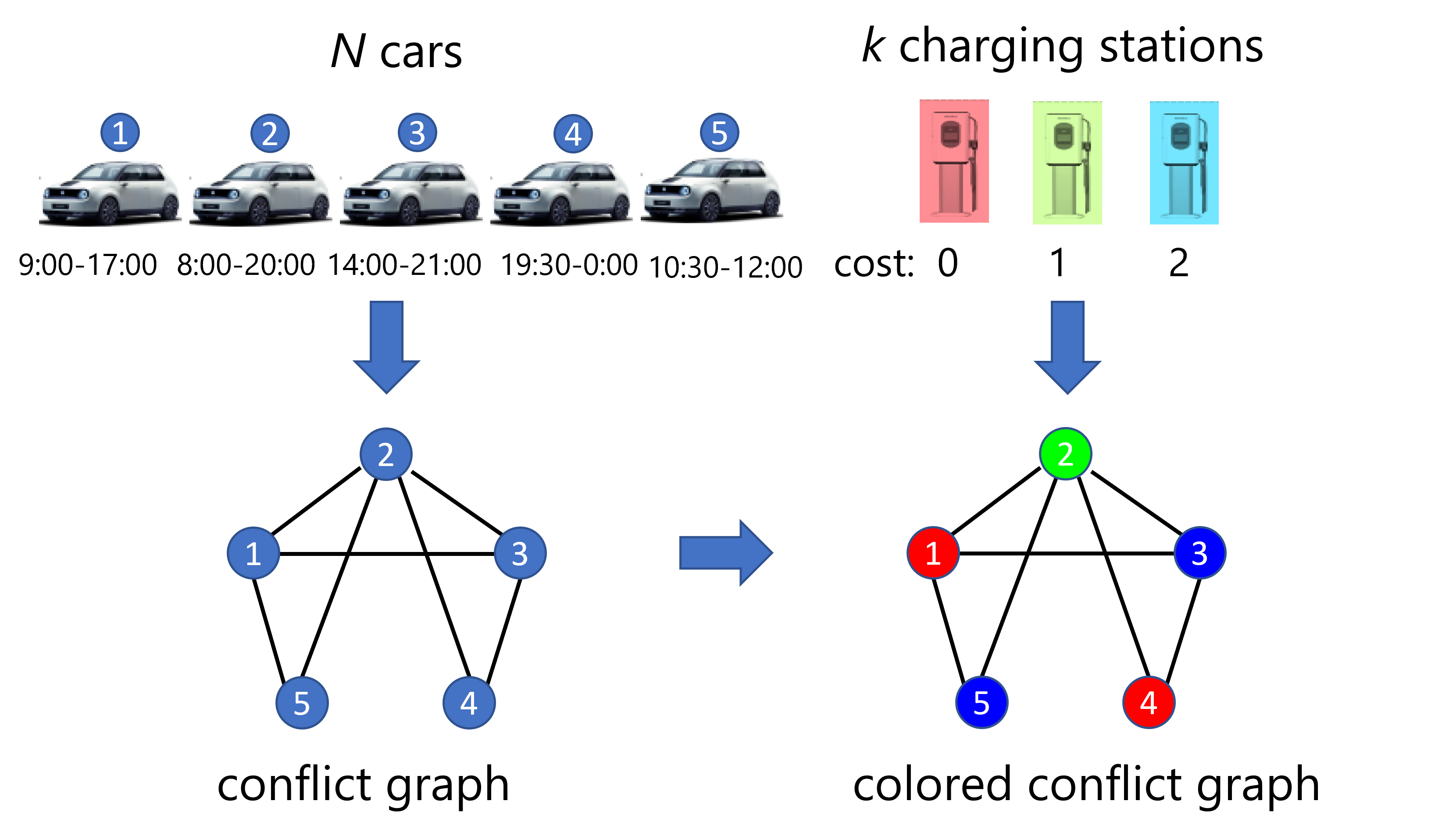} 
          \caption{\textbf{Schematic representation of the simplified EV charging problem.}
          $N$ cars need to be assigned to $k$ charging stations, where each car needs to be placed at a charging station for a given time period Here, we consider $N=5$ cars and $k=3$ charging stations.  
          No two cars with overlapping charging periods can be assigned to the same charging station which can be formulated as a conflict graph where cars with overlapping time slots are connected by an edge. If we denote each charging station
          by a different color, the charging station assignment can be formulated as a coloring problem of the conflict graph.
          Further, we assume that each charging station incurs  different costs, which are dimensionless numbers. 
       \label{fig:EVcharging}
        }
\end{figure}

Here we consider the EV charging optimization problem of the previous subsection with additional constraints on the charging time slots. Therefore we consider a graph where each vertex represents an EV, and each edge indicates overlapping charging time slots. Furthermore, the vertex color represents the number of the charging station, whereby each charging station has different costs. Finally, the constraint that two EVs cannot be charged at the same station simultaneously is modeled by the condition that two connected vertices must not have the same color. This charging problem is schematically illustrated in Fig.~\ref{fig:EVcharging}.

A cost function, which combines the different charging station costs and penalizes charging two vehicles at the same station simultaneously, is 
\begin{align} \label{eq:CostFunctionCombined}
C(\mathbf{z})= \sum_{n=1}^N\sum_{i=0}^{k-1} c_i z_n^i+\lambda \sum_{(u, v) \in E} \mathrm{\delta}_{z_{u}, z_{v}}\,,
\end{align} 
where  $z_n$ is the color of node $n$, the parameters $c_i$ encode  the cost for each color, and $\lambda>0$ is the penalty factor. 
Note that we always use dimensionless cost functions and thus the $c_i$ and the penalty factors  $\lambda$ are also dimensionless.
When all colors have equal costs, all coefficients $c_i$ for $i>0$ are zero and this problem reduces to the pure max-$k$-coloring problem of Sec.~\ref{sec:graphColor}.
For the case with three colors, $k=3$, and associated costs $c_{-1}$, $c_0$ and $c_1$, the cost Hamiltonian is 
\begin{align} \label{eq:CostHamiltonianExtended}
&H_C= \sum_{n=1}^N\left(c_0+ \tfrac{c_1-c_{-1}}2 L_{z,n} + \tfrac{c_1+c_{-1}-2c_0}2 L_{z,n}^2\right) \notag \\
&+ \lambda\!\!\!\!\sum_{(n, m) \in E} \bigg[1-L_{z,n}^2-L_{z,m}^2+\frac{1}{2}L_{z,n}L_{z,m} 
+\frac{3}{2}L_{z,n}^2L_{z,m}^2\bigg],
\end{align}
where the vertices are denoted by $n$ and $m$. Specifically, the coloring constraint induces two-site interactions with up to quadratic terms in $L_z$ on each vertex.

\section{Numerical Results}\label{PerformanceAnalysis}
In the following, we discuss numerical results of the QAOA for the cost Hamiltonian Eq.~\eqref{eq:CostHamiltonianExtended} for three colors $k=3$. We consider different graphs ranging from $N=4$ to $N=9$ and include the graph-coloring constraint term with a penalty factor $\lambda=20$ directly in the quantum circuit, as detailed in section \ref{sec:eqConstraint}. Further, we consider two cases for the costs: (i) $(c_{-1},c_0,c_1)=(0,0,0)$, where the charging problem reduces to  max-$k$-graph coloring, and (ii) $(c_{-1},c_0,c_1)=(0,1,2)$. A recent work~\cite{bravyi2020}  benchmarked QAOA on pure max-$k$-graph coloring with qutrits ($k=3$) on random 3-colorable constant-degree graphs up to a size of $N=300$, which was possible for $p=1$ (a circuit with one layer). In contrast, we employ several layers ranging from $p=1$ to $p=8$, introduce an additional cost contribution for each vertex color and focus on individual instances of highly connected graphs. Further, we compare the performance of two classical optimization algorithms for the cost function encoded by Eq.~\eqref{eq:CostHamiltonianExtended}. 

We employ two classical optimization algorithms: the limited-memory Broyden–Fletcher–Goldfarb–Shanno (L-BFGS) algorithm~\cite{scipy} and the covariance matrix adaptation evolutionary strategy (CMA-ES)~\cite{hansenCMA} taken from~\cite{cmaesRepo}. For each setting (graph, cost function, and circuit depth), we execute 50 different QAOA optimization runs of the CMA-ES with randomly chosen initial values for $\bm{\gamma}$ and $\bm{\beta}$. The population-based CMA-ES evaluates between 6 and 12 candidate solutions in each generation, depending on the search space dimension $2p$. For the L-BFGS optimizer, we use between 300 and 600 optimization runs to be comparable to the number of circuit evaluations with CMA-ES.

 \begin{figure}
    \centering
     \hspace*{-0.47\columnwidth}{(a)} \hspace*{0.45\columnwidth}{(b)}\\[-0mm]
     \includegraphics[width=0.23\textwidth]{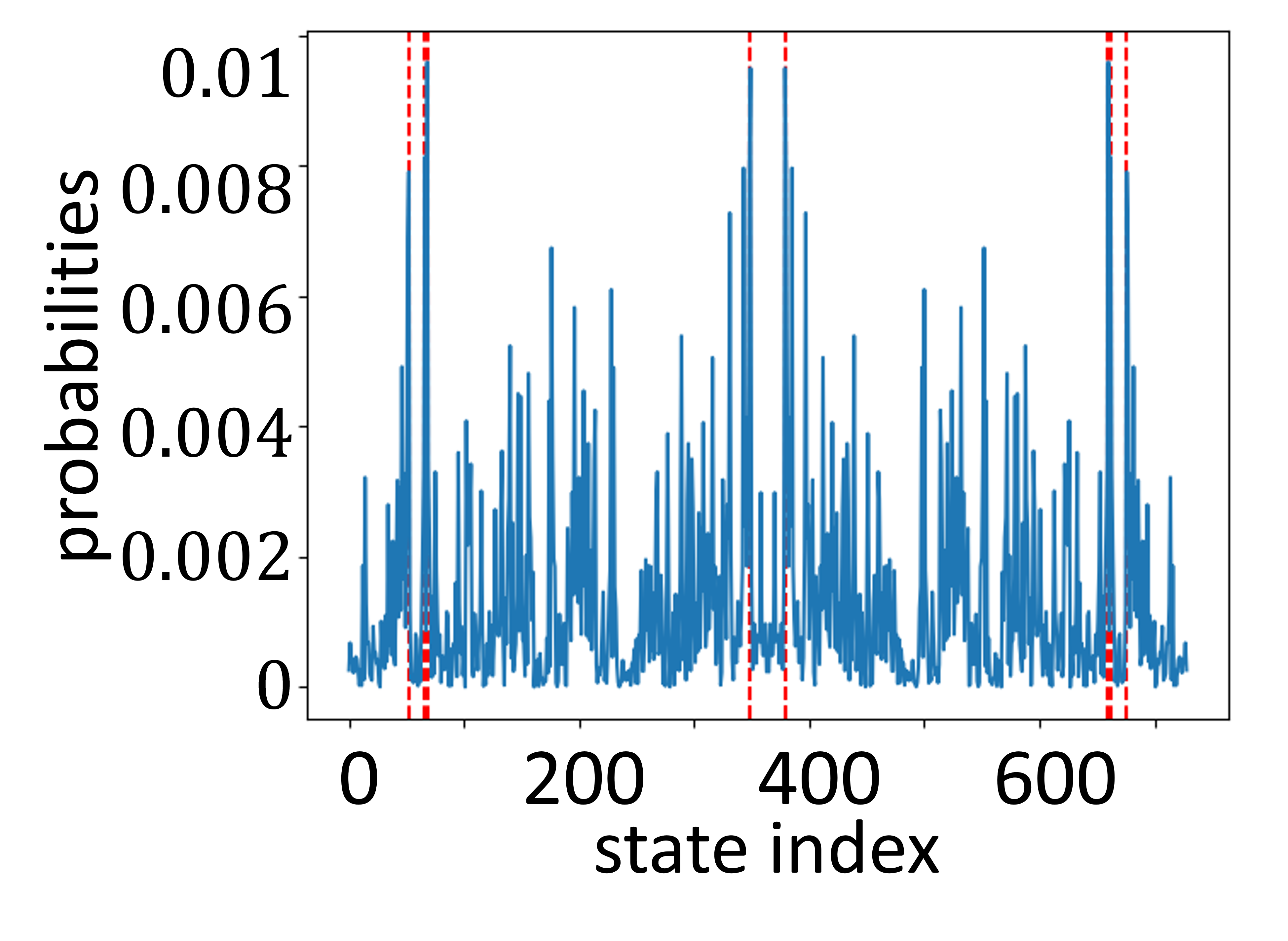} 
    \includegraphics[width=0.23\textwidth]{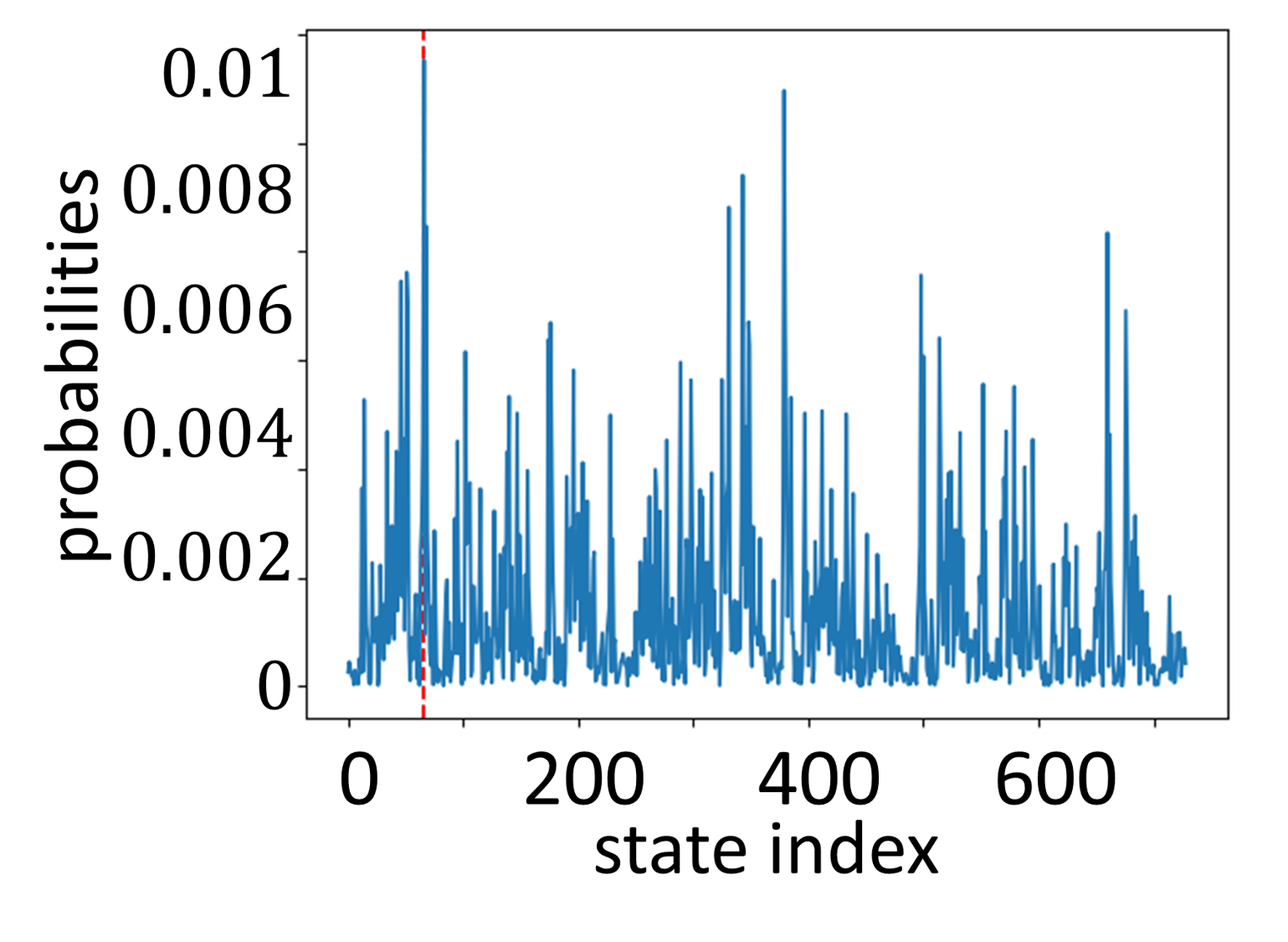}  
    \\
         \hspace*{-0.47\columnwidth}{(c)} \hspace*{0.45\columnwidth}{(d)}\\[-0mm]
     \includegraphics[width=0.23\textwidth]{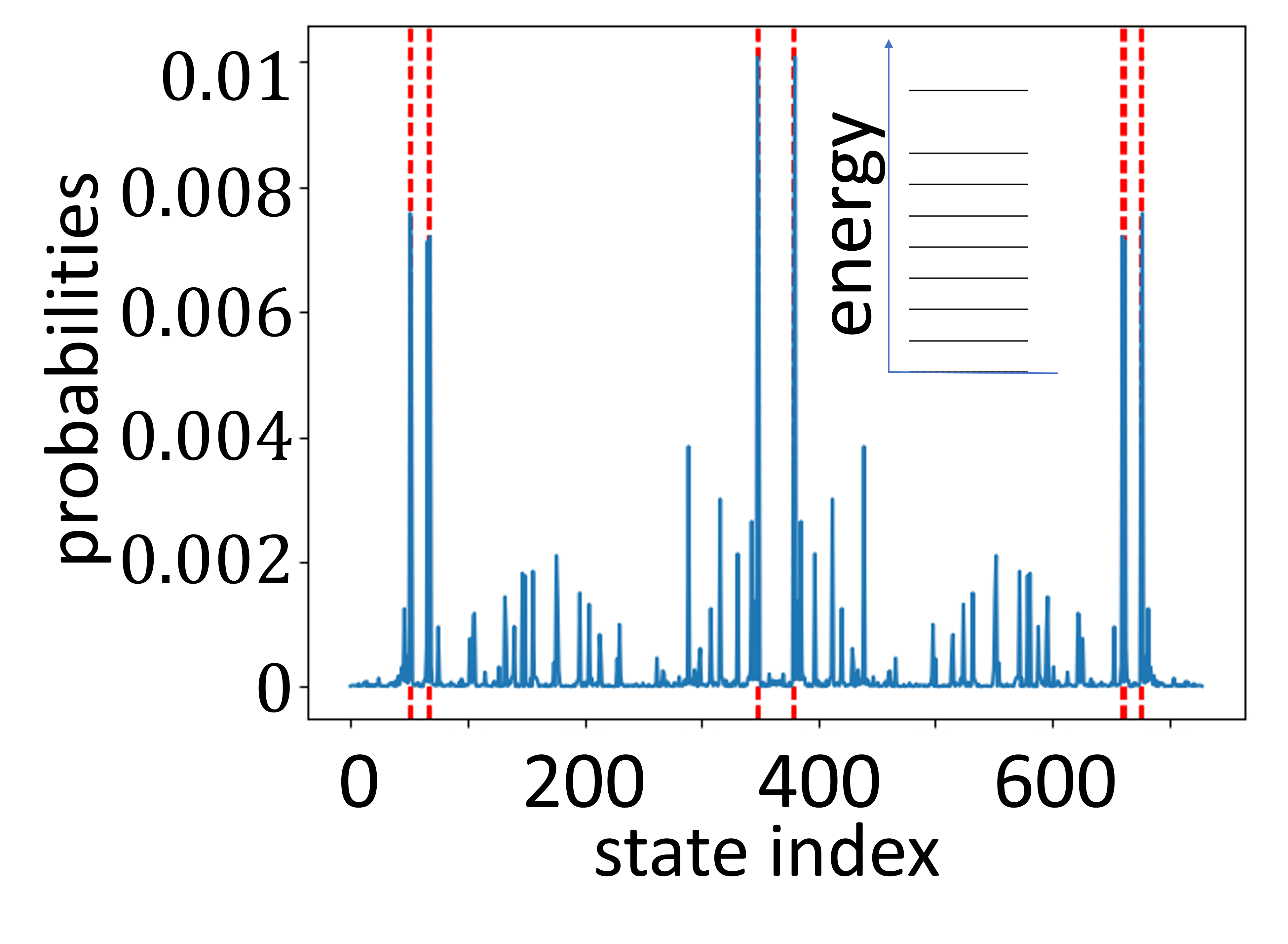}
    \includegraphics[width=0.23\textwidth]{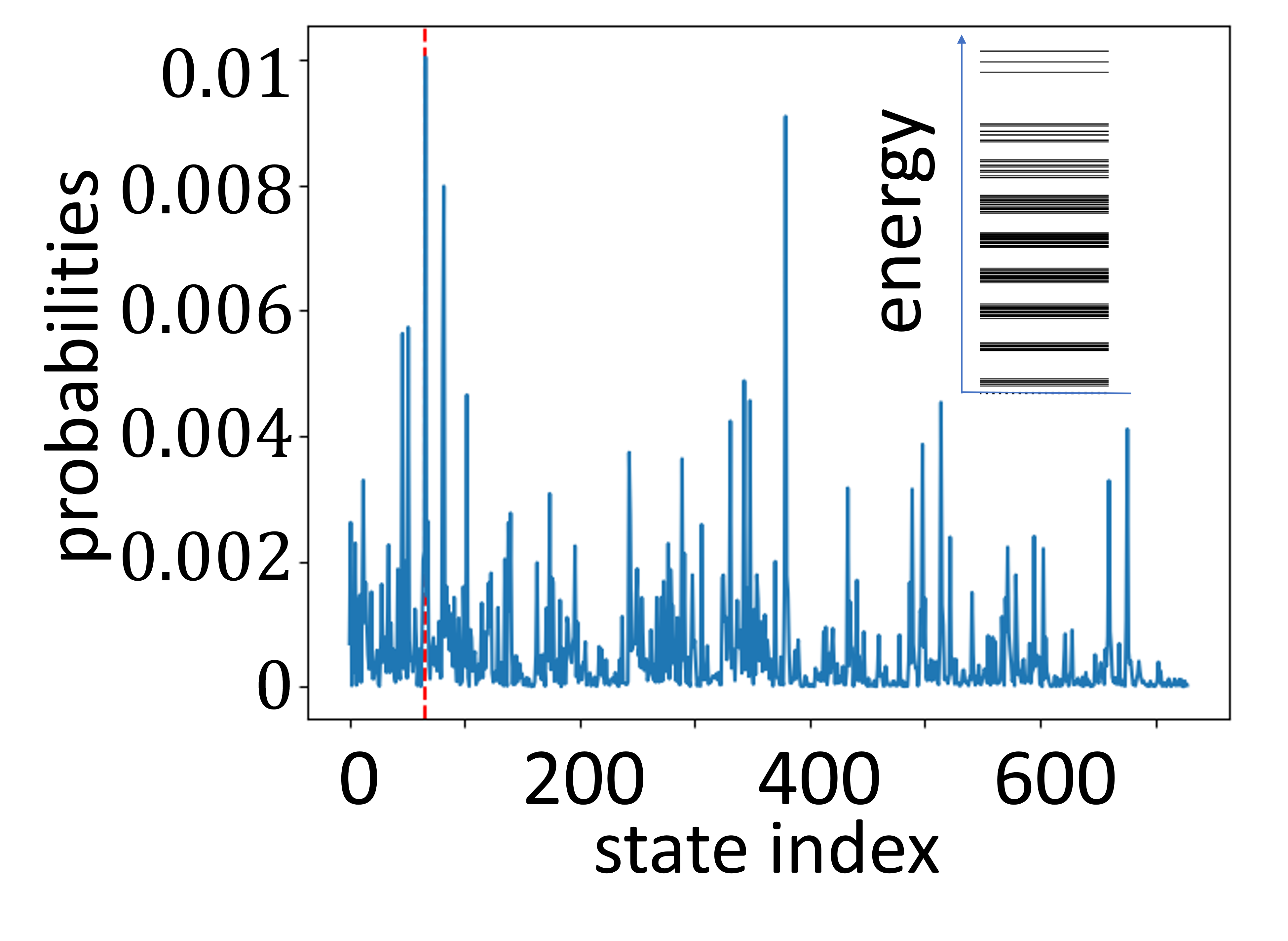}\\
    \hspace*{-0.47\columnwidth}{(e)} \hspace*{0.45\columnwidth}{ \phantom{(b)} }\\[-2mm]
     \includegraphics[width=0.11\textwidth]{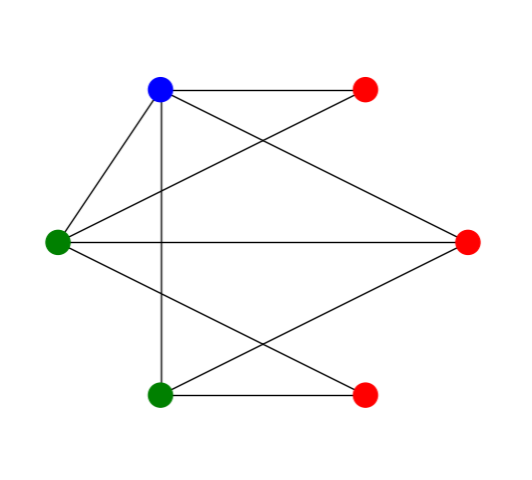}
    \includegraphics[width=0.11\textwidth]{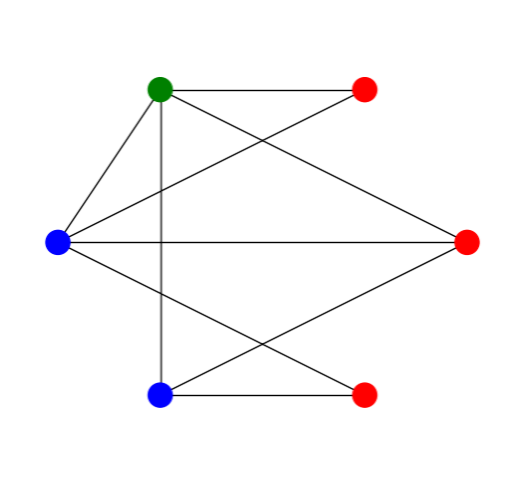}
    \includegraphics[width=0.11\textwidth]{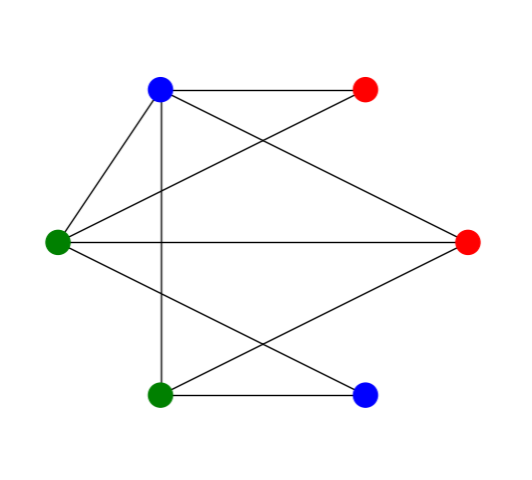}
    \includegraphics[width=0.11\textwidth]{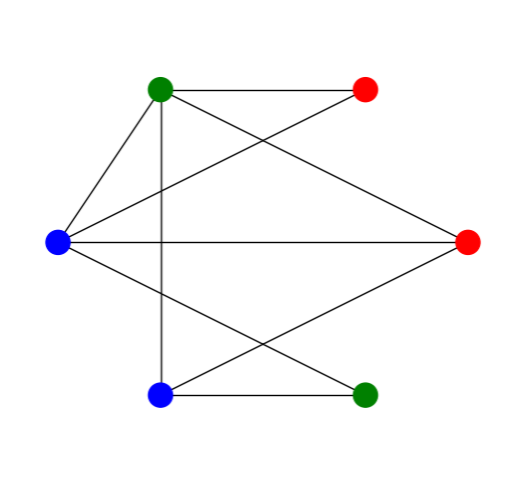}\\
          \caption{\textbf{Probability distributions of the optimized state and optimal solution graphs.}
       Probability distribution of representative final QAOA states for a $N=6$ graph without coloring cost, see (a) and (c) and with coloring cost $(c_{-1},c_{0},c_{1}) =(0,1,2)$, see (b) and (d). The upper row (a) and (b) depicts results for shallow circuits with depth $p=1$  while the middle row (c) and (d) shows results for depth $p=5$.  The red dashed lines indicate the 12 (1) optimal solutions without (with) coloring cost. The insets in panels (c) and (d) shows the energy spectra of the respective Hamiltonian in arbitrary units.
       (e) Candidate solutions for the simplified charging problem where the colors red, green, and blue have cost $c_{-1}$, $c_{0}$, and $c_{1}$, respectively. Without coloring cost all shown graphs are optimal solutions and the additional eight optimal solutions can be generated by pairwise color exchange. With coloring cost $(c_{-1},c_{0},c_{1})=(0,1,2)$, only the leftmost graph of panel (e) is optimal.       \label{fig:weights}
        }
\end{figure} 
Fig.~\ref{fig:weights} shows the probability distribution of the computational basis states determined by the final QAOA state for $N=6$, different circuit depth, and with/without coloring cost. For example, the pure graph coloring problem without coloring cost has twelve optimal states for the graph given in Fig.~\ref{fig:weights}. The probability distribution of the computational basis states is shown in panels (a) and (c), where example graph colorings are depicted in panel (e). However, lifting the color symmetry by including costs for different colors leads to a single optimal solution, as can be seen from the asymmetric probability distribution in panels (b) and (d). 

 \begin{figure}[th]
    \centering
     \hspace*{-0.47\columnwidth}{(a)} \hspace*{0.45\columnwidth}{(b)}\\[-3mm]
     \includegraphics[width=0.23\textwidth]{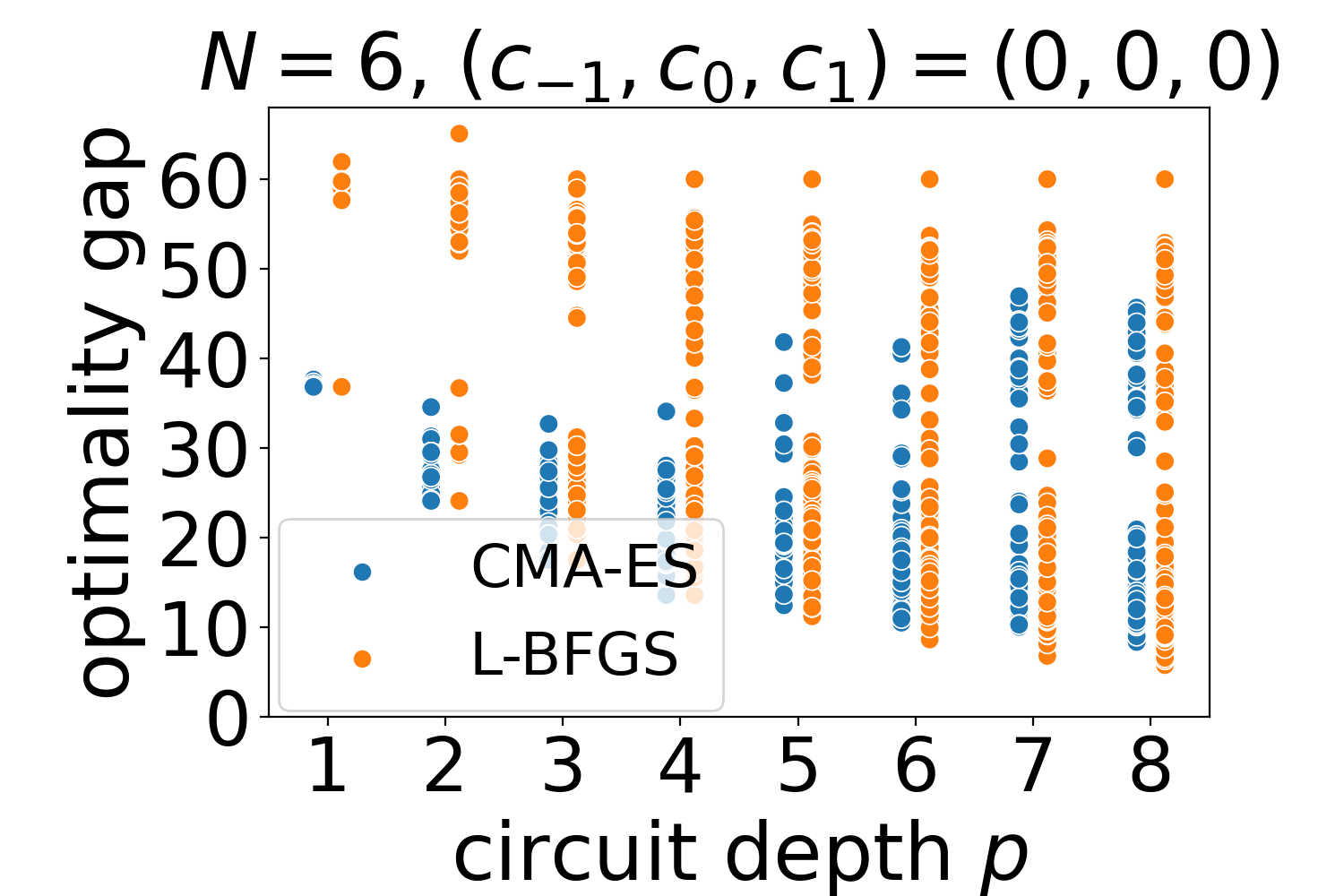}
     \includegraphics[width=0.23\textwidth]{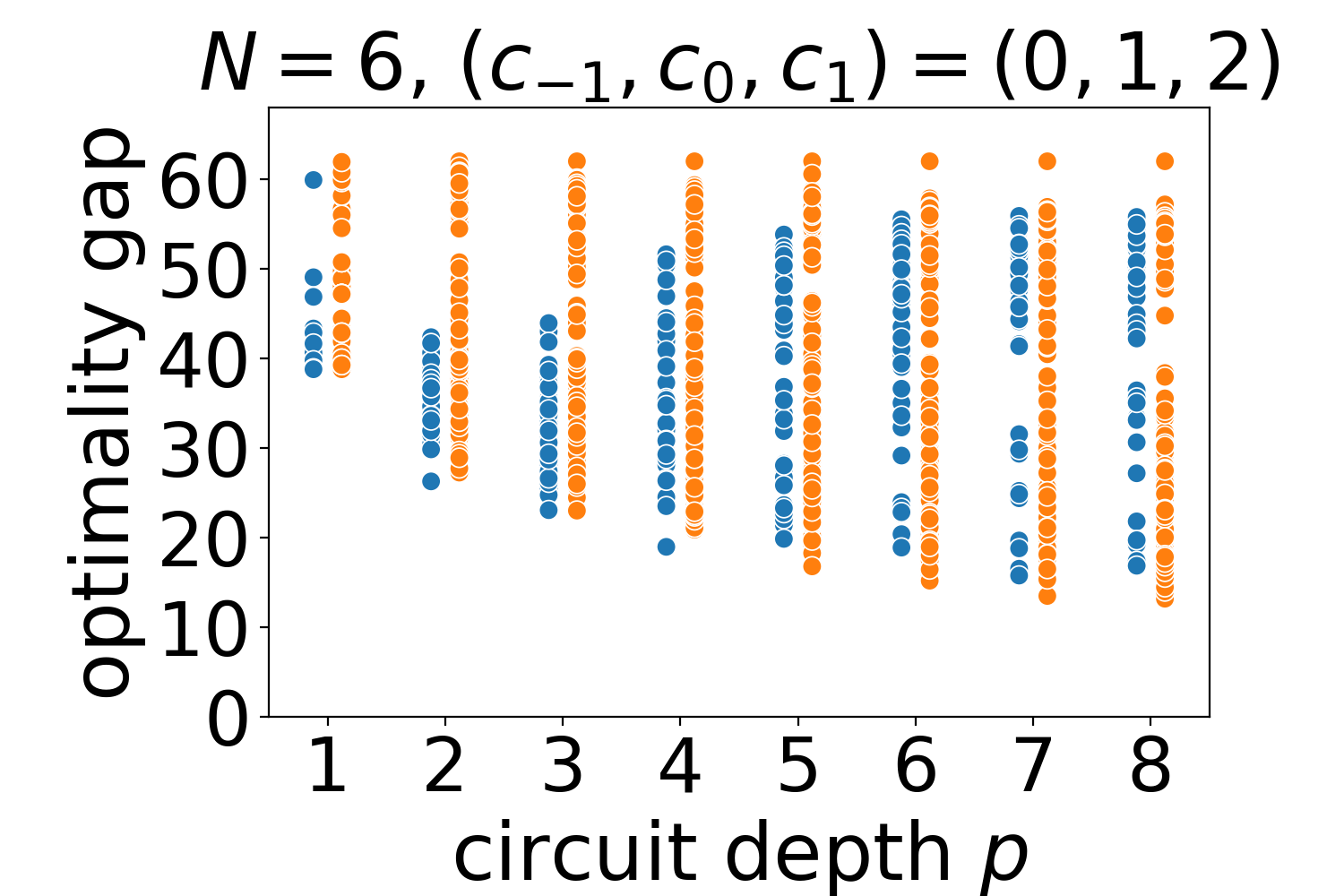}\\
     \hspace*{-0.47\columnwidth}{(c)} \hspace*{0.45\columnwidth}{(d)}\\[-0.5mm]
     \includegraphics[width=0.23\textwidth]{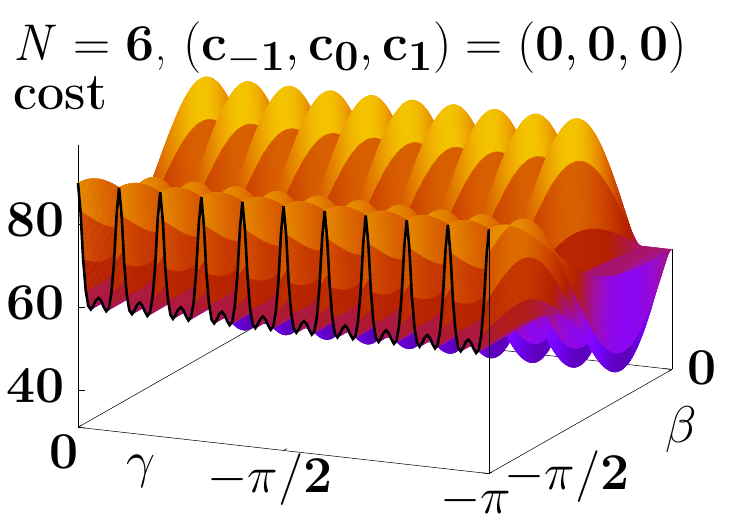}
     \includegraphics[width=0.23\textwidth]{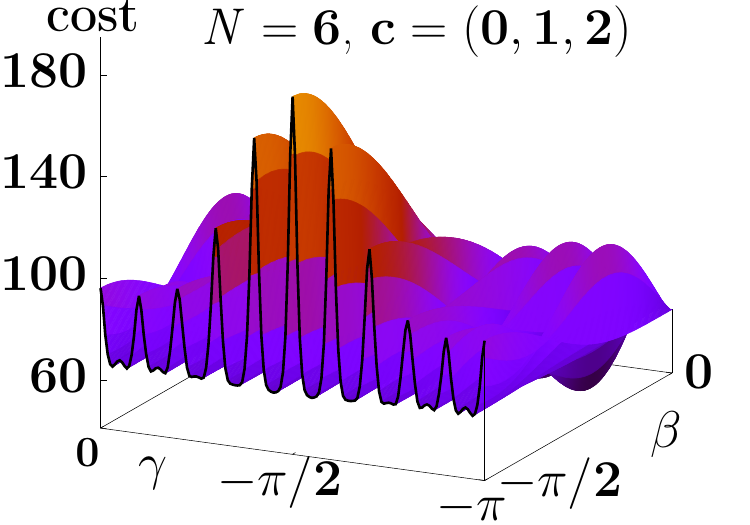}
     \caption{\textbf{Optimality gap and cost function landscape.}
       Optimality gap of the final QAOA states for a graph with $N=6$ (a) without coloring cost, i.e., $(c_{-1},c_{0},c_{1})=(0,0,0)$, and (b) with coloring cost $(c_{-1},c_{0},c_{1})=(0,1,2)$, as a function of the circuit depth $p$ and for the two optimizers L-BFGS (orange) and CMA-ES (blue). The plots show the best result for each of the 50 (300) CMA-ES (L-BFGS) runs. 
       Panels (c) and (d) show the cost function landscape for circuit depth $p=1$ in a certain parameter region of the search space for the same $N=6$ graph without (c) and with (d)  coloring cost.}
        \label{fig:N6_fitness}
\end{figure} 

In  Fig.~\ref{fig:N6_fitness}, we show results for a representative example of the simplified charging problem on a graph with $N=6$ nodes and three colors ($k=3$) with and without coloring cost. Fig.~\ref{fig:N6_fitness}a and Fig.~\ref{fig:N6_fitness}b show the optimality gap, i.e., the difference between the exact minimum and the minimum obtained from the QAOA cost function for different circuit depths $p\leq 8$. The exact minimum was obtained by exhaustive search of the whole state space, which was possible for the limited problem size considered here. Generally, the lowest value of the optimality gap decreases for both optimizers with increasing circuit depth, indicating that a deeper circuit can, in principle, achieve smaller values of the cost function~\cite{zhou2020,Moussa2020}. However, there are considerable variations in the optimality gap between the runs.

This behavior of the optimality gap is understandable as the cost function landscape of the QAOA is typically highly multi-modal with many local minima and maxima, as can be seen in Fig.~\ref{fig:N6_fitness}c and Fig.~\ref{fig:N6_fitness}d where a part of the  $p=1$  two-dimensional  search landscape is shown. The multi-modal and in particular the ridge-like structure of the cost function landscape makes the optimization problem considerably harder for algorithms like L-BFGS which use gradient information, as it introduces saddle-point like features known to cause problems in many settings including deep-learning applications~\cite{dauphin14,Choromanska2015,BottouDNNReview2018}. Hence, the L-BFGS optimization process may end up in a local minimum with a high probability,  essentially determined by the location of the random initial starting point. In contrast, the CMA-ES is a population-based global optimizer capable of dealing with this cost function landscape and finds lower cost minima more reliably.  
 
\begin{figure}
    \centering
     \hspace*{-0.47\columnwidth}{(a)} \hspace*{0.45\columnwidth}{(b)}\\
     \includegraphics[width=0.23\textwidth]{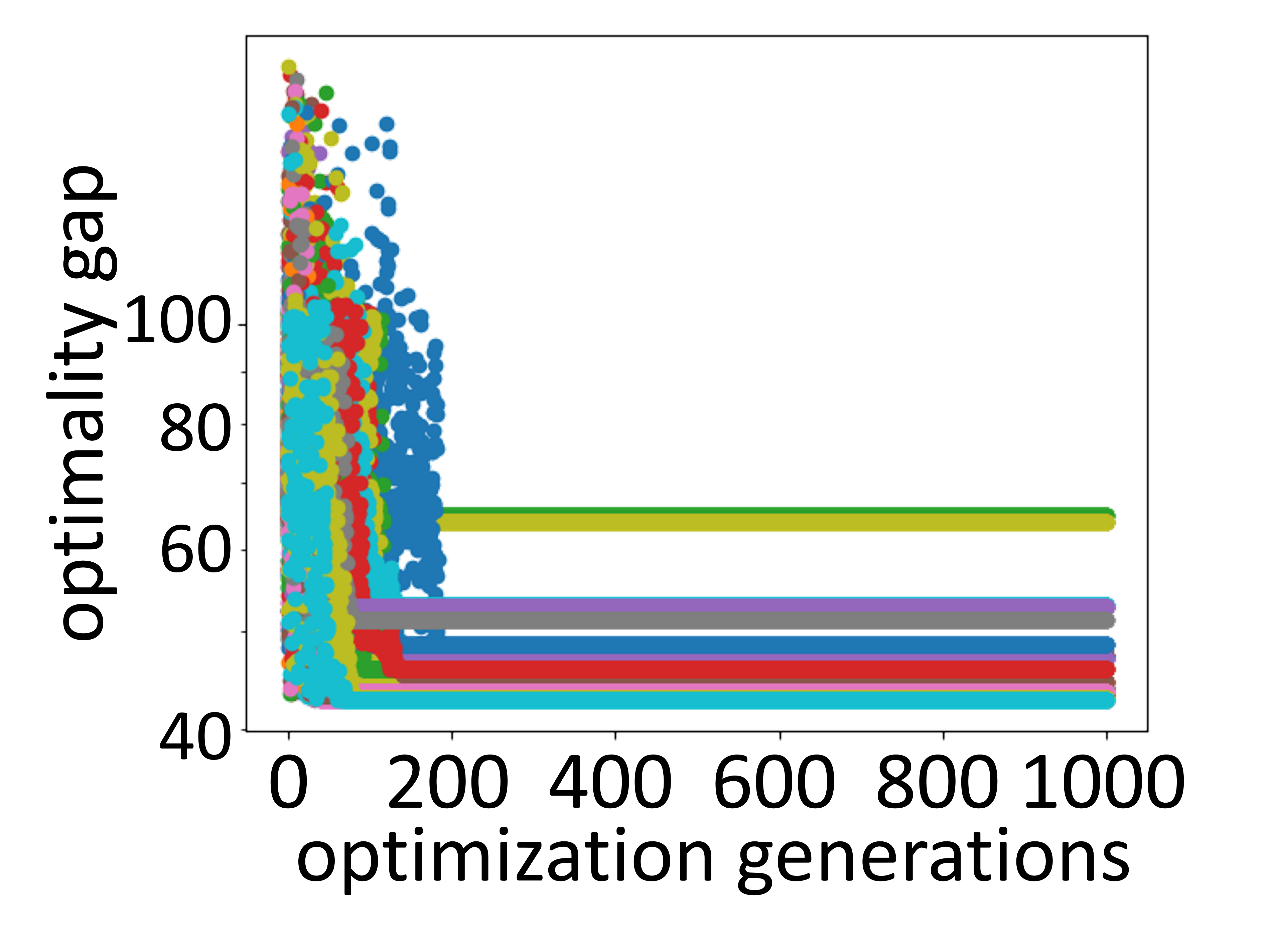}
     \includegraphics[width=0.23\textwidth]{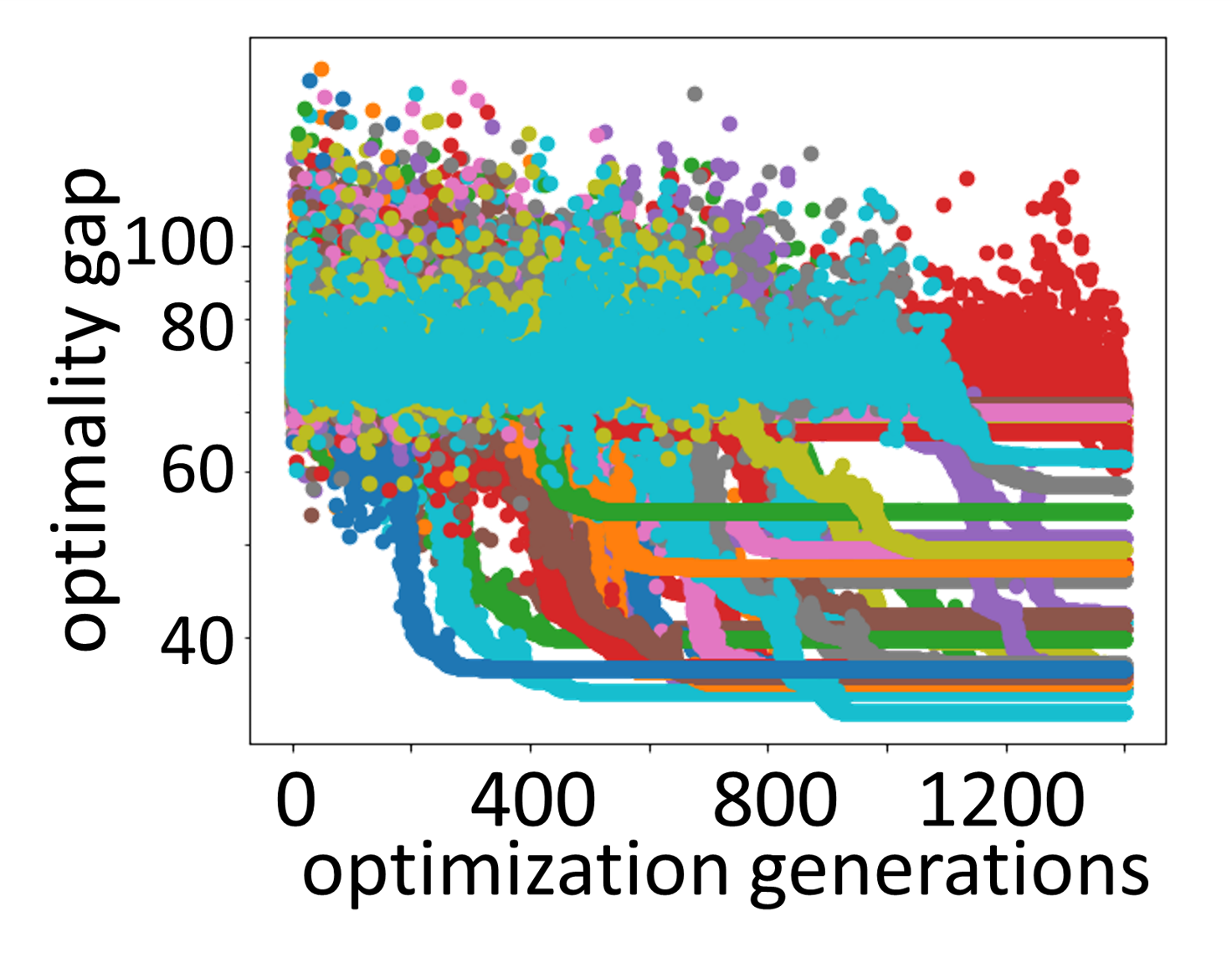}
      \caption{\textbf{Optimization progress.} Optimality gap values of $50$ CMA-ES optimization runs as function of the internal optimization generation number for a $N=6$ graph without coloring cost and with a circuit depth of $p=1$ (a) and $p=5$ (b). Each color represents one individual optimization run with different initial values for the search parameters $\bm{\gamma}$ and $\bm{\beta}$        }
        \label{fig:CMAESprogress}
\end{figure} 

However, for larger circuits, particularly for $p\gtrsim4$, the CMA-ES may also result in sub-optimal local minima, a common behavior for evolutionary algorithms in a larger search space. In Fig.~\ref{fig:CMAESprogress} we show the progress of the CMA-ES optimization runs as a function of the internal generation number. For the more straightforward problem with $p=1$ shown in panel (a), the CMA-ES converges rather quickly in about 200 generations. In contrast, for the more challenging optimization problem with $p=5$, we can observe that some runs do not converge even after 1400 generations. In principle, this limitation can be removed by running the algorithms for more generations, which comes at the cost of more circuit evaluations.

It is instructive to investigate how the improvement of the optimality gap for deeper circuits (c.f.\ Fig.~\ref{fig:N6_fitness}) is reflected in the probability distribution. For shallow circuits, i.e., $p=1$, the global minima are clearly visible, as indicated by the red dashed lines in Fig.~\ref{fig:weights}a and Fig.~\ref{fig:weights}b. However, other states with higher energy do have sizable contributions in the probability distribution of the final state. Increasing the circuit depth leads to better separation of the optimal states compared to sub-optimal states and a smaller value of the cost function, see panels Fig.~\ref{fig:weights}c and Fig.~\ref{fig:weights}d .

Comparing the probability distribution of the case without color cost in Fig.~\ref{fig:weights}a and c and with color cost in Fig.~\ref{fig:weights}b and d shows a qualitative difference, which can be understood from the spectrum of the cost Hamiltonian. The spectrum of the pure graph coloring Hamiltonian, see the inset of Fig~\ref{fig:weights}c, has a large gap between the (degenerate) ground state manifold and the first excited states. As the QAOA circuit is an approximation to an adiabatic time evolution, a large energy gap between the ground state and excited state is beneficial for finding the ground state. In contrast, we do not observe a clear energy gap when considering coloring costs, see the inset of Fig~\ref{fig:weights}d. Especially the appearance of multiple low-lying energy states makes the problem of separating those states in the quantum circuit harder. Consequently, non-optimal states have higher amplitudes for the problem containing coloring costs compared to the pure graph coloring problem.

 \begin{figure}
    \centering
    \hspace*{-0.47\columnwidth}{(a)} \hspace*{0.45\columnwidth}{(b) }\\[-2mm]
     \includegraphics[width=0.23\textwidth]{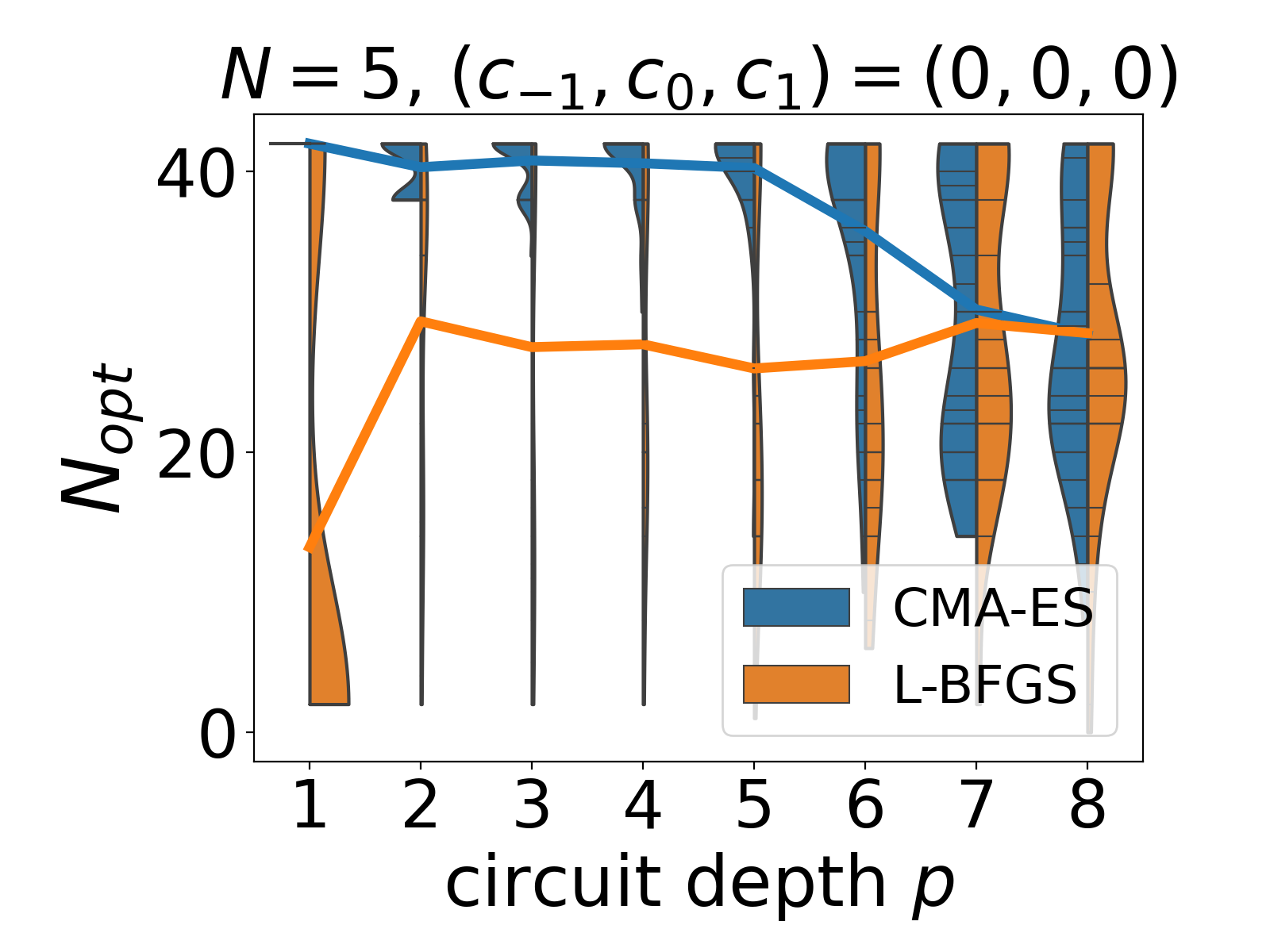}
     \includegraphics[width=0.23\textwidth]{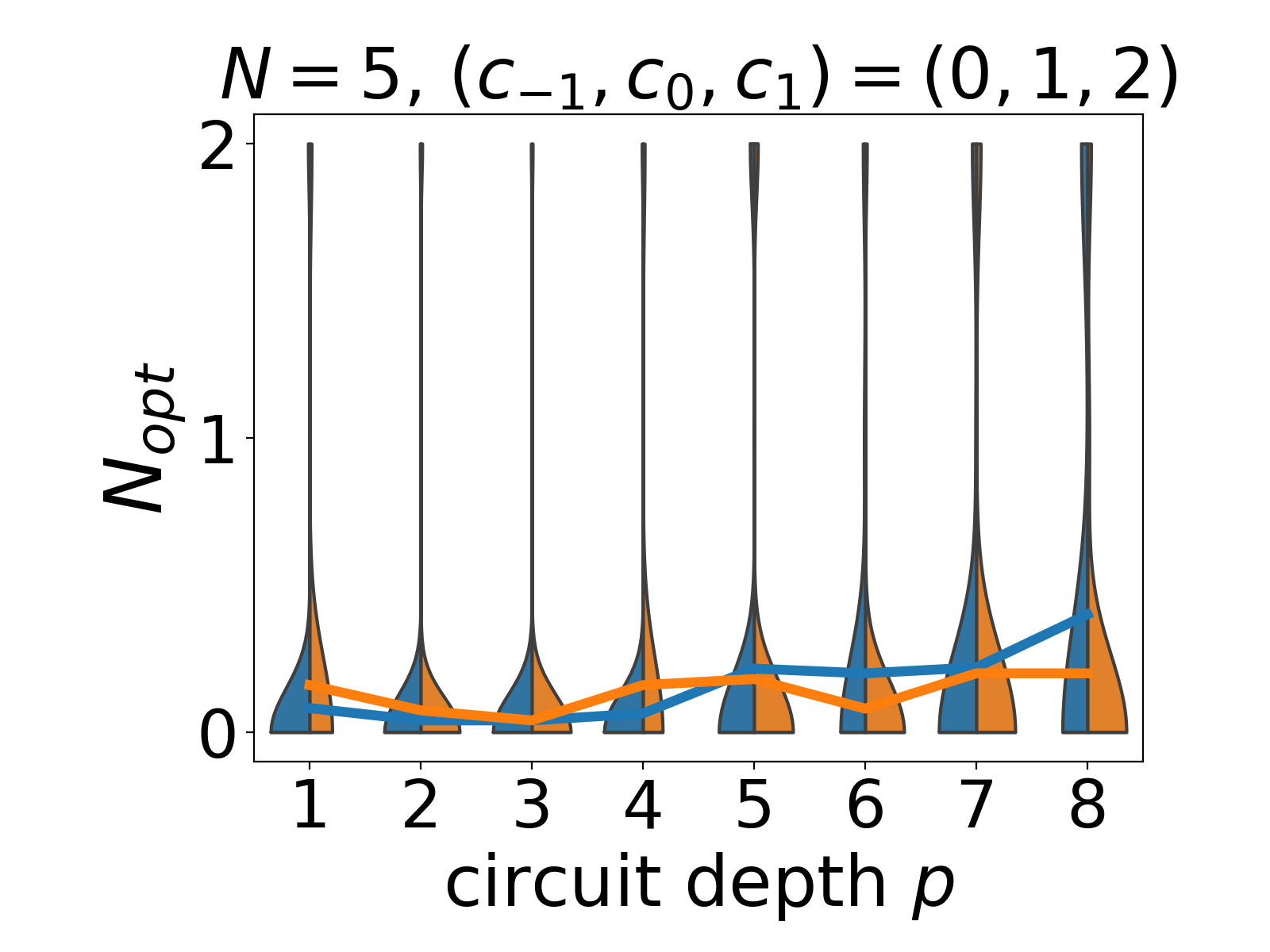} \\ 
         \hspace*{-0.47\columnwidth}{(c)} \hspace*{0.45\columnwidth}{(d) }\\[-2mm]
     \includegraphics[width=0.23\textwidth]{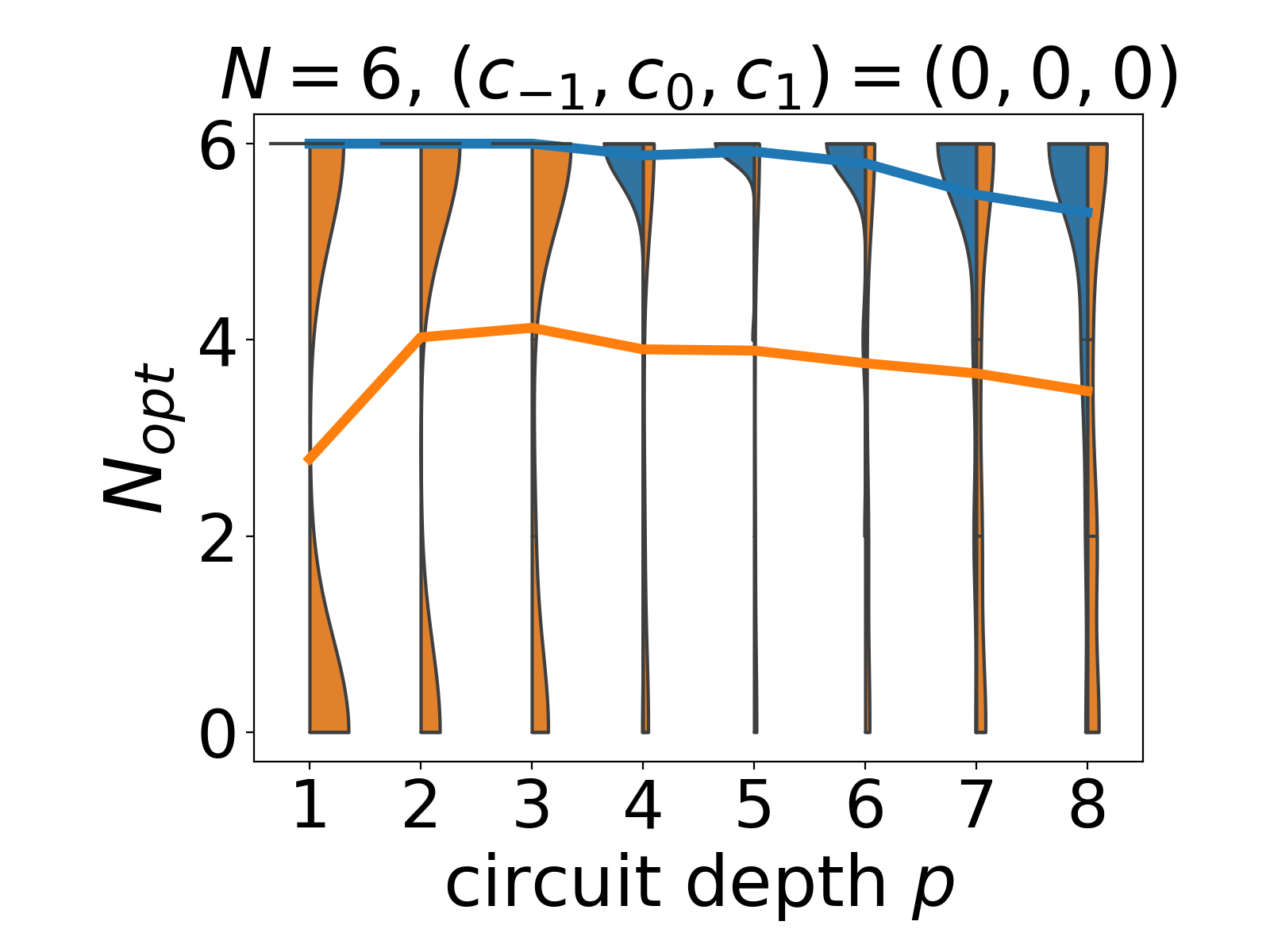}
     \includegraphics[width=0.23\textwidth]{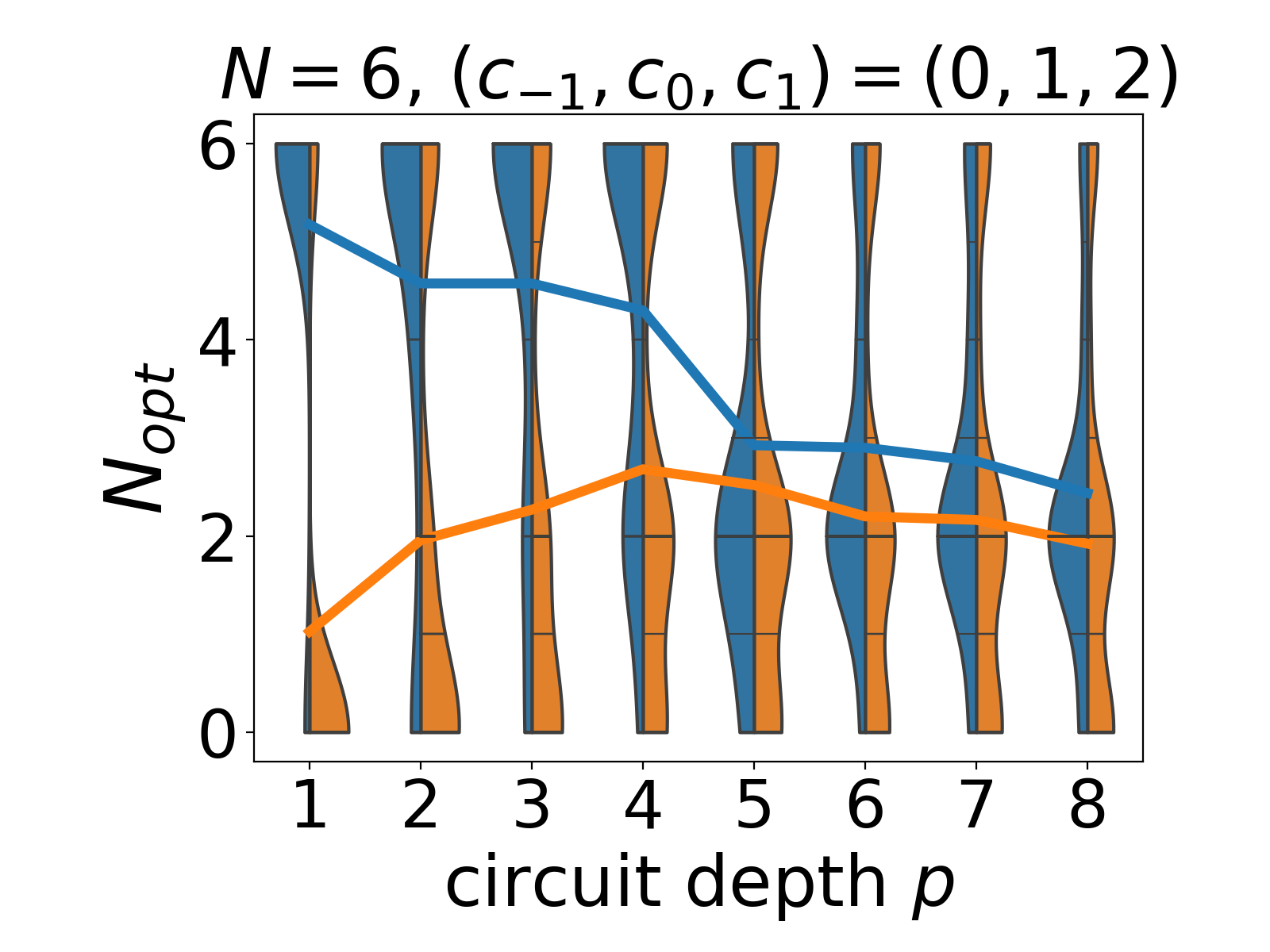}\\ 
         \hspace*{-0.47\columnwidth}{(e)} \hspace*{0.45\columnwidth}{(f) }\\[-2mm]
     \includegraphics[width=0.23\textwidth]{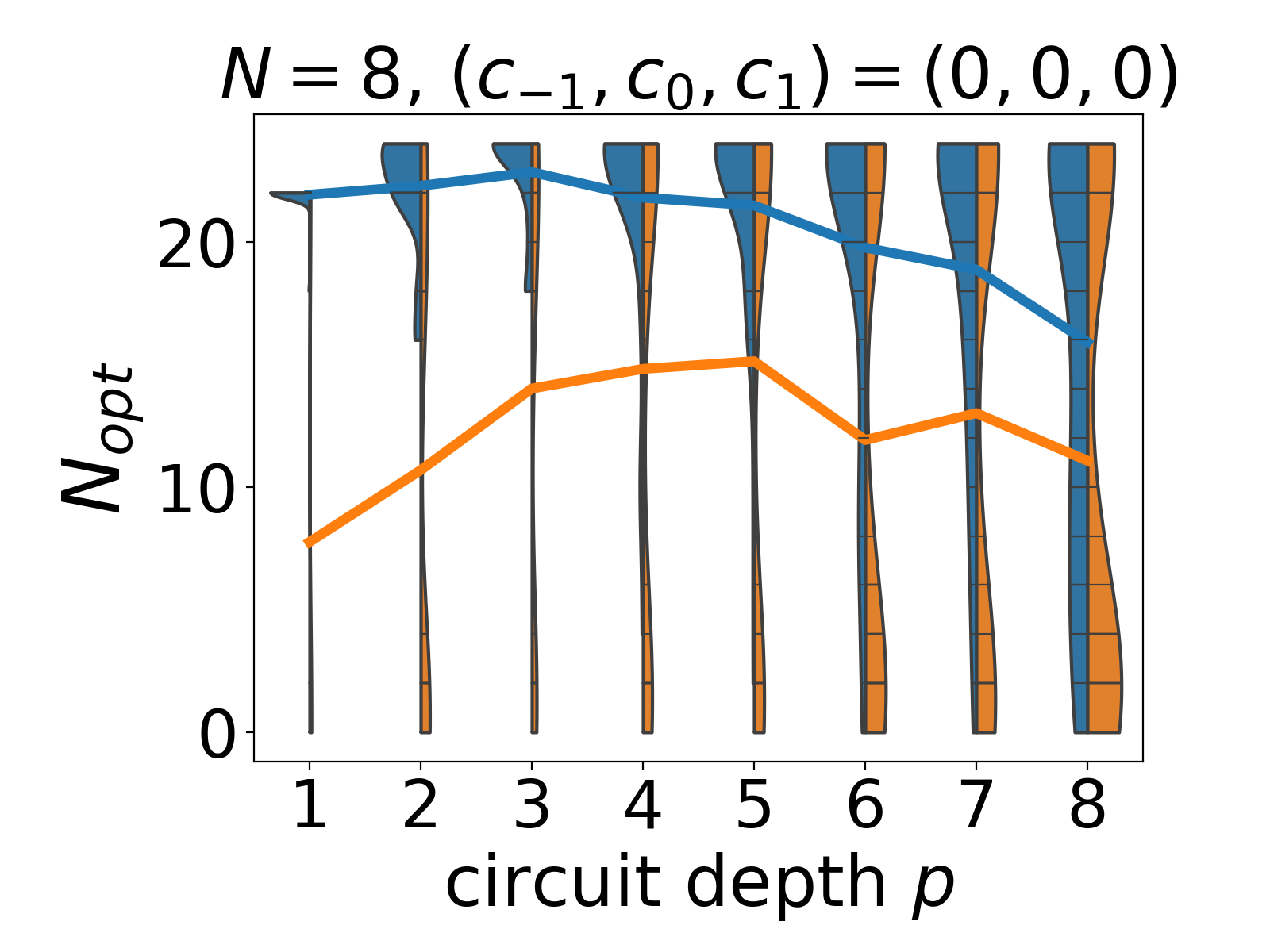}
     \includegraphics[width=0.23\textwidth]{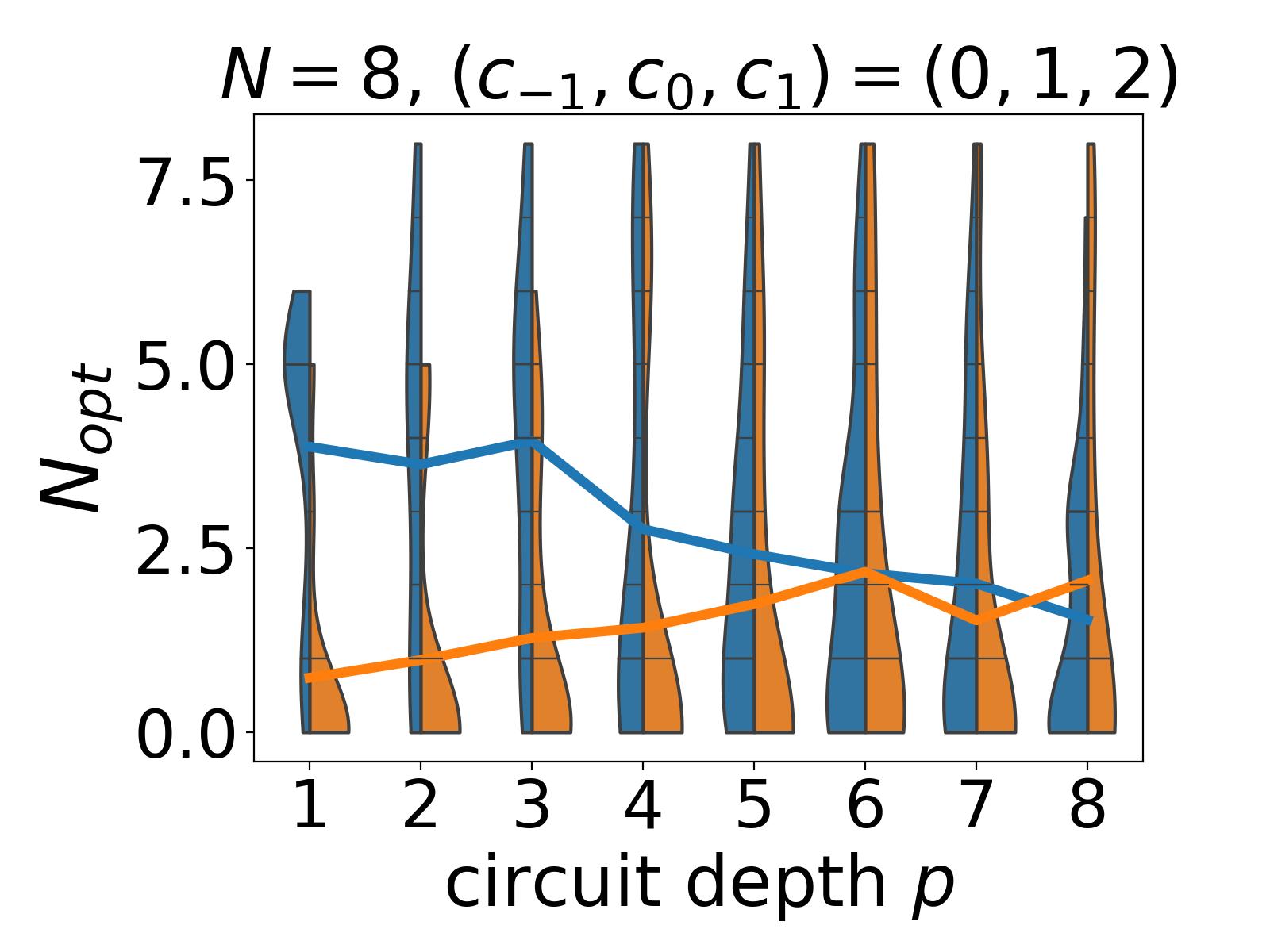}
      \caption{\textbf{Distribution of optimal solutions over several QAOA runs.}
       Number of optimal solutions found in one QAOA optimization run for simplified charging problem instances on graphs with $N=5$ (panels (a) and (b)), $N=6$ [(c) and (d)] and $N=8$ [(e) and (f)] without [(a), (c), and (d)] and with charging cost. The colored bars show the number of found optimal solutions in each run, where larger width implies larger number of found solutions. 
       Lines show the mean number of found solutions aggregated over all  $50$ and $300-600$ different optimization runs using CMA-ES and L-BFGS, respectively. 
        }
        \label{fig:Nopt}
\end{figure} 

By preparing and sampling from the final QAOA state, one can then extract multiple candidates for the optimum of the cost function. We extract several candidate solutions and test for optimality for different graphs for the simplified charging problem with and without coloring cost. In addition, we perform several optimization runs and depict the average number of optimal solutions (with and without coloring costs) in Fig.~\ref{fig:Nopt}. The vertical axis of Fig.~\ref{fig:Nopt} always extends to the total number of optimal solutions (determined by a classical optimizer), and the error bars indicate the minimal and maximal numbers of solutions found with the QAOA after the classical optimization loop. 
In most cases, we can find $\boldsymbol{\beta}$ and $\boldsymbol{\gamma}$ such that the QAOA state allows for detecting all optimal solutions regardless of the circuit depth, the optimization algorithm, and the problem instance. Exceptions, where the QAOA could not find all optimal solutions are shown in Fig.~\ref{fig:Nopt}f for $N=8$ and $p\leq3$ with coloring costs, and for $N=8$ and $p=1$ without coloring cost.  

Inspecting the average number of found optimal solutions, we can observe that the variation between different runs is substantial. In particular, for the instance without coloring cost on a graph with $N=5$ nodes shown in Fig.~\ref{fig:Nopt}(a), the QAOA reliably finds almost all 42 optimal solutions when employing the CMA-ES for not too deep circuits, while introducing coloring cost on the same graphs, see Fig.~\ref{fig:Nopt}(b), leads to a situation where on average almost none of the two optimal solutions are found. We point out
that finding all optimal solutions with the CMA-ES becomes
more difficult with increasing circuit depth, which
is a typical behavior for evolutionary algorithms when increasing 
parameter space. Both optimization algorithms show lower performance in finding all optimal solutions in one run on average for the problem with color cost than the pure graph coloring problem. This effect results from lifting the degeneracy by introducing the coloring cost, which leads to multiple low-lying states close in energy. The
relation between the spectrum of the Hamiltonian and the performance of the 
QAOA is a field of current study~\cite{zhou2020}.

\section{Conclusion}
In this article, we discussed the QAOA for current or upcoming qudit experiments. First, we described how to map cost functions onto cost Hamiltonians utilizing generalized Pauli or angular momentum operators. Additionally, we illustrated different ways to incorporate equality or inequality constraints. Therefore, we laid out a scheme to include constraints into the classical optimization loop. We also presented three alternative methods to incorporate constraints into the quantum circuit. The first method adds penalty functions for the constraints into the cost Hamiltonian. The second method realizes equality and inequality constraints using conditional gates and ancilla qubits, similar to an error-correcting code. Finally, the third approach implements constraints by exploiting dynamical decoupling, which suppresses computational basis states which violate equality constraints. 

As an application of the QAOA with qudits, we discussed theoretical and industry-relevant optimization problems, for example, the graph $k$-coloring  or an EV charging problem with global power constraints. Since these problems only involve bounded integer variables, they can be mapped on qudits. Motivated by current experiments with ultracold atoms or ions, we propose to use the $x$-angular momentum operator $L_x$ as a mixing operator. Finally, we numerically studied a simplified EV charging optimization problem, which amounts to a max-$k$-graph coloring problem with an additional color cost term on the vertices. 
We compared solutions of the QAOA obtained with the gradient-based classical L-BFGS optimizer and the global evolutionary CMA-ES optimizer for our numerical studies. Our results showed that the global evolutionary optimizer was less sensitive to the initialization of the search and reliably produced better results than the gradient-based approach for the instances considered. This performance behavior can be understood by the highly multi-modal cost function landscape. Starting from pure max-$k$-coloring and introducing a coloring cost function, the performance with both optimizers typically decreased. This behavior is a direct consequence of reducing the symmetry of the cost Hamiltonian. 

We extracted solutions from the final state by selecting states with the largest amplitudes. In our examples, we could find multiple optimal solutions. In particular, the final state is also symmetric when the cost Hamiltonian is invariant with respect to a symmetry, and the mixing operator does not break this symmetry. The final state thus includes equal-weight superpositions of symmetry-related states. Finding several candidate solutions is a highly desirable feature for practical applications, as one has the chance to obtain a large subset of all possible solutions. However, these degeneracies may lead to detrimental performance since the amplitude of the optimal states may be distributed such that the sampling of the candidate solutions becomes inefficient. Notably, the signal-to-noise ratio between optimal and sub-optimal states may be reduced. In order to improve the signal-to-noise ratio, one can single out optimal states by investigating ways to reduce the number of candidate solutions, e.g., via sparsity constraint, on the QAOA trial state. This approach is left for future study.

In this work, we studied a selection of optimization problems. However, we expect that the insights generated here are relevant for general problem instances on larger graphs and different types of problems. Specifically, the cost function landscape will generally be multi-modal due to the structural form of the mixer and phase separation operators. Therefore global black-box optimizers are expected to be very useful for the QAOA~\cite{lavrijsen2020}. 
Another promising and highly relevant aspect of qudit-based implementations of QAOA is the possibility of resource-efficient implementation on hardware, which was shown in previous work~\cite{Weggemans2021Solving}. The question how this advantage over qubit-based implementations extends to the formulation of realistic problems including constraints in details  is left for future research. 
In total, we have extended the QAOA toolbox for qudit systems and applied it to relevant theoretical and practical applications opening up the avenue for current and future qudit platforms to solve integer optimization problems. 
  
\section{Acknowledgements}
We acknowledge fruitful discussions with R.~Blatt, A.~Bottarelli, A.~Garcia-Sala, D.~Gonzalez-Cuadra, M.K.~Oberthaler, M.~Ringbauer, H.~T\"ureci, T.V.~Zache,  and P.~Zoller.

S. L.\ acknowledges support from project Quantum Hub Th\"uringen, 2021 FGI 0047, Free State of Thuringia, Th\"uringer Aufbaubank.

F.J.\ acknowledges the DFG support through the project FOR
2724, the Emmy- Noether grant (Project-ID 377616843).
This work is  supported by the DFG Collaborative Research Centre "SFB 1225 (ISOQUANT)", by the Bundesministerium f\"ur Wirtschaft und Energie through the project "EnerQuant" (Project- ID 03EI1025C) and the Bundesministerium f\"ur Bildung und Forschung through the project "HFAK" (Project- ID 13N15632).

P.H.\ acknowledges support by Provincia Autonoma di Trento, the ERC Starting Grant StrEnQTh (project ID 804305), the Google Research Scholar Award ProGauge, and Q@TN — Quantum Science and Technology in Trento.  

V.K.\ and M.L.\ acknowledge support from: ERC AdG NOQIA; Agencia Estatal de Investigaci\'{o}n (R\&D project CEX2019-000910-S, funded by MCIN/ AEI/10.13039/501100011033, Plan National FIDEUA PID2019-106901GB-I00, FPI, QUANTERA MAQS PCI2019-111828-2, Proyectos de I+D+I “Retos Colaboración” QUSPIN RTC2019-007196-7); Fundaci\'{o} Cellex; Fundaci\'{o} Mir-Puig; Generalitat de Catalunya through the CERCA program, AGAUR Grant No. 2017 SGR 134, QuantumCAT \ U16-011424, co-funded by ERDF Operational Program of Catalonia 2014-2020; EU Horizon 2020 FET-OPEN OPTOLogic (Grant No 899794); National Science Centre, Poland (Symfonia Grant No. 2016/20/W/ST4/00314); Marie Sk\l odowska-Curie grant STREDCH No 101029393; "La Caixa" Junior Leaders fellowships (ID100010434) and EU Horizon 2020 under Marie Sk\l odowska-Curie grant agreement No 847648 (LCF/BQ/PI19/11690013, LCF/BQ/PI20/11760031,  LCF/BQ/PR20/11770012, LCF/BQ/PR21/11840013).

\appendix

\section{Realization of the qudit-QAOA with atomic systems}\label{SpinPhononSystems}
This appendix discusses the experimental capabilities of ultracold atoms to realize the angular momentum encoding of quadratic cost functions and the mixing Hamiltonian given in Eq.~\eqref{eq:StandardMixer}. Specifically, quadratic cost functions can be experimentally realized in three distinct atomic platforms: cold atomic mixtures~\cite{Kasper2020}, cold quantum gases in a cavity and Rydberg atoms \cite{CavitySystem, PRXQuantum.2.020319}. 
In both systems, the qudit is realized as a long collective spin by cooling atoms with internal degrees of freedom into the ground state of optical lattice sites. In the mixture system, the effective interaction between different qudits is mediated by phononic excitations, theoretically proposed in \cite{Kasper2020}. In the cavity system, the long-range interaction between the atoms is mediated via a photonic mode, which was experimentally demonstrated with high control over the interaction and the connectivity in Ref.~\cite{CavitySystem}. 

Both the mixture and the cavity system are described by the effective Hamiltonian 
\begin{align} \label{QuadraticHamiltonian}
    H_{C} = \sum_{\mathbf{x},\mathbf{y}} U(\mathbf{x},\mathbf{y}) L_z(\mathbf{x}) L_z(\mathbf{y}) + \sum_{\mathbf{x}} b(\mathbf{x}) L_z(\mathbf{x}),
\end{align}
where $\mathbf{x}$ and $\mathbf{y}$ denote the minima of the lattice potential, $U(\mathbf{x},\mathbf{y})$ is the long-range potential between the qudits, and $b(\mathbf{x})$ is a locally controllable 
energy shift. The mixing Hamiltonian can be engineered by standard tools such as global microwave pulses~\cite{FisherInfo} that lead to terms of the form
\begin{align}
    H_M = \Omega \sum_{\mathbf{z}}  L_x(\mathbf{x})\,.
\end{align}
A major advantage of employing these two platforms with high connectivity is the natural implementation of quadratic cost functions. 

However, quadratic Hamiltonians do not suffice to encode the cost functions for all problems we consider in this work, e.g., the graph coloring of Sec.~\ref{sec:graphColor}. Nevertheless, cost functions containing higher powers 
of angular momentum operators may be implemented by Trotterization or by employing resource Hamiltonians, as demonstrated in the context of variational quantum simulation~\cite{Kokail2019}. Another possibility 
is to employ a universal quantum computer which is based on qudits.
For example, trapped ion platforms are able to implement generalized Pauli operators and can entangle qudits and as such can implement the QAOA, see Ref.~\cite{IonQudits2021} for more details.

\section{Other optimization problems}
\label{appendix}
This appendix introduces several optimization problems whose cost functions are naturally expressed in terms of qudits, namely the knapsack problem, multiway number partitioning, and job-shop scheduling. 
 
 \subsection{Knapsack problem}
The knapsack problem consists in assigning a set of items to a container~\cite{Lucas2014}. There are $N$ different items with $c$ copies each. Further, each item $i$ has a weight $w_i$ and a value $v_i$, and the goal is to maximize the total value in the container while not exceeding a given weight limit $W$. The cost function of the bounded knapsack problem~\cite{Knapsack1995} is 
 \begin{align}
 C(\bm{z})&=\sum_{i=1}^{N} v_{i} z_{i} \, ,
 \end{align}
which has to be maximized and is subject to the weight constraint 
 \begin{align}
 \label{eq:knapConstr}
 \sum_{i=1}^{N} w_{i} z_{i} \leq W\,,
 \end{align}
 where $z_{i} \in[0,c]$.
 
The bounded knapsack problem can be straightforwardly mapped to qudits by using the angular momentum encoding discussed in Sec.~\ref{Encoding} promoting the integer variables $z_i$ to qudits with $d = c+1$. Using the angular momentum operator $L_{z,i}$ for $\ell = c/2$ we obtain the cost function
\begin{equation}
     H_C = -\sum_{i=1}^{N} v_{i} L_{z,i}\,,
\end{equation}
where we included a minus sign in order to transform the problem into minimization problem. The constraints~\eqref{eq:knapConstr} are linear in the angular momentum operators and can be implemented by using the methods developed in Sec.~\ref{Constraints}. 

 \subsection{Multiway number partitioning }
 The number partitioning problem is the task of partitioning a list $S$ of $n$ positive integers, $S=\left( s_{1}, s_{2}, \ldots, s_{n}\right)$, into $k$ subsets $S_{1}, S_{2}, \cdots, S_{k}$, such that the numbers are as equally distributed as possible. That is, the sum of the numbers in different subsets $V_{i}=\sum_{l \in S_{i}} s_{l}$ for $1 \leq i \leq k$ is requested to be as similar as possible. For instance, if $S=(1,1,2,3,4,5)$ and $k=2$, the optimal partitions are $(1,1,2,4)$ and $(3,5)$, which in this case yields a completely balanced partition with $V_1=V_2=8$. 
 For the case of $k=2$ the decision version of number-partitioning problem is NP-complete~\cite{Mertens2003}, though there are various algorithms that solve the problem efficiently in many cases. 
A trapped-ion setup for two-way number partitioning has been proposed in Ref.~\cite{Hauke2015}. 

The multi-way number partitioning can be cast into a mathematical cost function as follows. 
The sum of the elements in the set $S_{i}$ is
 \begin{align}
    \label{eq:Viz}
     V_i(\mathbf{z}) = \sum_{l=1}^n s_l \delta_{i,z_l} \,, 
 \end{align}
 where the value of the variable $z_l=1,\dots,k$ indicates the subset $S_i$ of which $s_l$ is a member. 
 We then choose the cost function
 \begin{equation}
     C(\mathbf{z}) = \sum_{a<b} [V_a(\mathbf{z}) - V_b(\mathbf{z})]^2\,,
 \end{equation}
 which minimizes the differences between the sum of the partitions with the vector $\mathbf{z}=(z_1,\dots,z_n)$. 
 
  The implementation in $k$-level systems $z_l$ automatically ensures that each $s_l$ is member of exactly one set $S_i$ with $i=1,\dots,k$. 
  In general, the realization of $\delta_{i,z_l}$ in Eq.~\eqref{eq:Viz} requires a polynomial of order $k$ in the $z_l$. For example, for $k=3$, $\delta_{1,z_l}=3-5z_l/2 + z_l^2/2$. Importantly, these higher-order terms within $V_i(\mathbf{z})$ act locally, but the qudits are then coupled in a pair-wise fashion via $C(\mathbf{z})$. 
  
 In the literature there exist various other approaches to mathematically formulate the multiway number partitioning problem, which become equivalent in the case of $k=2$, see~\cite{korf2010objective}. 
 Here, we have opted for a cost function that employs integer variables and leads to a direct construction using only two-qudit interactions.

\subsection{Job-shop scheduling}
\label{sec:jss}
 \begin{figure}[ht]
      \centering
      \includegraphics[width=\columnwidth]{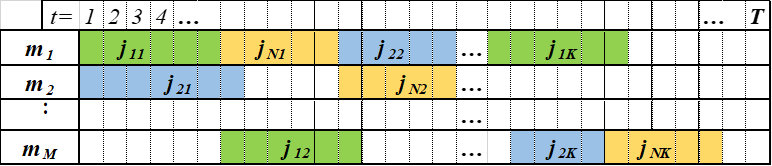}   
      \caption{\textbf{Job-shop scheduling.} Table representation of a job schedule. The horizontal line denotes the discretized time, 
      whereas the vertical axis denote the machine. Filling the box 
      corresponds to using the machine with the job $j_{n,k}$.}
     \label{fig:jobshop}
 \end{figure}

 The problem consists of the task to schedule the execution of $N$ jobs $j$ on $M$ machines. 
 Each job is subdivided into $K$ operations, where $j_{n,k}$ denotes the operation $k$ of job $n$, and each operation has a predefined processing duration $p_{n,k}$, where $n\in[1,N]$ and $k\in[1,K]$.
 The operations of one job must be executed in a predefined order $ j_{n,1}\rightarrow j_{n,2}\rightarrow\dots \rightarrow j_{n,K}$ and must not overlap. 
 Further, each operation $j_{n,k}$ has to be executed on one specific machine $m_{n,k}\in[1,M]$ and operations executed on one machine must not overlap. A schematic representation of this problem is shown in Fig.~\ref{fig:jobshop}.

 An encoding based on qudits is formulated by discretizing the time into $T$ equally space time intervals, $t=1,\dots,T$.
 The problem is then formulated with  the  variables $t_{n,k}\in\{1,\dots,T\}$ which specify the time at which the execution of operation  $j_{n,k}$ on machine $m_{n,k}$ starts.  

 There are two constraints to be respected. First,  two operations of the same job must not overlap, i.e., the predecessor operation must finish before the successor can start:
 \begin{align}
 \label{eq:JSS_D2}
 & t_{n,k}+p_{n,k} < t_{n,k+1} \, ,
 \end{align}
 which needs to to be fulfilled for all $k\in [1,K-1]$ and all jobs $n$.
 Second, two operations on one machine must not overlap, i.e., only one operation can run at any given time on one machine. 
 This means that for any two operations  $j_{n,k}$ and $j_{n',k'}$ to be executed  on the same machine, the one operation must either be finished before the other operation or start after it, i.e.,
 \begin{align}
 \label{eq:JSS_C3}
 & (t_{n,k}+p_{n,k} < t_{n',k'} ) \,\text{XOR}\, (t_{n,k} >  t_{n',k'}+p_{n',k'}) 
 \end{align}
 for all machines $m$ and all $ (n,k),(n',k') \in o_m$ where $o_m=\{(n,k) | m_{n,k}=m\}$ is the list of operations to be run on machine $m$.
 The latter condition can also be transformed into a quadratic constraint 
 \begin{align}
 \label{eq:JSS_C3a}
  (t_{n,k}+p_{n,k} - t_{n',k'} ) (t_{n,k} -  t_{n',k'}-p_{n',k'}) >0\,.
 \end{align}

 Depending on the application scenario, multiple different cost functions can be employed \cite{JainJSSReview1999, kuMIP_JSS2016}. 
 A typical cost function is given by the average job completion time, 
 \begin{align}
     C(\bm{t})&=\frac1N\sum_{n=1}^N(t_{n,K}+p_{n,K})\, ,
 \end{align}
 which represents overall machine-usage efficiency and needs to be minimized.

Another cost function is the makespan, which is the finishing time of the last operation,
 \begin{align}
  C(\bm{t})&=\max_n(t_{n,K}+p_{n,K})\,.
  \end{align}
 In order to avoid the nonlinear max-function, a linear cost function can be formulated with an additional auxiliary variable,  
 \begin{align}
  C'(\bm{t}')&=t_{N+1,K}\,,
  \end{align}
  which needs to fulfill $N$ additional linear constraints
  \begin{align}
  t_{n,K}+p_{n,K}< t_{N+1,K}
  \end{align}
  for all jobs $n=1,\dots,N$.

 The above formulation directly lends itself to addressing the job-shop scheduling problem with QAOA by replacing the classical variable $t_{n,k}$ with an angular momentum operator with total spin $\ell=(T-1)/2$. 
 All constraints and the cost functions can be expressed as operators by replacing the classical variables with these $L_z$-operators.  
 Thus, we need  $kN$ qudits ($kN+1$ qudits) for the average completion time (makespan) formulation, where each qudit has dimension $d=T$.

%\bibliography{ref}
\input{main.bbl}

\end{document}